\documentclass[aps,prx,twocolumn,superscriptaddress]{revtex4-2}
\usepackage[utf8]{inputenc}
\usepackage[version=4]{mhchem}
\usepackage{multirow}
\usepackage{graphicx}
\usepackage{physics}
\usepackage{textgreek}
\usepackage{gensymb}
\usepackage{xcolor}
\usepackage[colorlinks]{hyperref}
\usepackage[capitalize]{cleveref}
\usepackage[caption=false]{subfig}

\newcommand{\supp}{Supplemental Material}

\begin{document}

\title{Fractional AC Josephson Effect in a Topological Insulator Proximitized by a Self-Formed Superconductor}
\author{Ilan T. Rosen}
\affiliation{Department of Applied Physics, Stanford University, Stanford, California 94305, USA}
\affiliation{Stanford Institute for Materials and Energy Sciences, SLAC National Accelerator Laboratory, Menlo Park, California 94025, USA}
\author{Christie J. Trimble}
\affiliation{Joint Quantum Institute and Quantum Materials Center, Department of Physics, University of Maryland, College Park, MD, USA}
\author{Molly P. Andersen}
\affiliation{Department of Materials Science and Engineering, Stanford University, Stanford, California 94305, USA}
\affiliation{Stanford Institute for Materials and Energy Sciences, SLAC National Accelerator Laboratory, Menlo Park, California 94025, USA}
\author{Evgeny Mikheev}
\affiliation{Department of Physics, Stanford University, Stanford, California 94305, USA}
\affiliation{Stanford Institute for Materials and Energy Sciences, SLAC National Accelerator Laboratory, Menlo Park, California 94025, USA}
\author{Yanbin Li}
\affiliation{Department of Materials Science and Engineering, Stanford University, Stanford, California 94305, USA}
\author{Yunzhi Liu}
\affiliation{Department of Materials Science and Engineering, Stanford University, Stanford, California 94305, USA}
\author{Lixuan Tai}
\affiliation{Department of Electrical Engineering, University of California, Los Angeles, California 90095, USA}
\author{Peng Zhang}
\affiliation{Department of Electrical Engineering, University of California, Los Angeles, California 90095, USA}
\author{Kang L. Wang}
\affiliation{Department of Electrical Engineering, University of California, Los Angeles, California 90095, USA}
\author{Yi Cui}
\affiliation{Department of Materials Science and Engineering, Stanford University, Stanford, California 94305, USA}
\affiliation{Stanford Institute for Materials and Energy Sciences, SLAC National Accelerator Laboratory, Menlo Park, California 94025, USA}
\author{M. A. Kastner}
\affiliation{Department of Physics, Stanford University, Stanford, California 94305, USA}
\affiliation{Stanford Institute for Materials and Energy Sciences, SLAC National Accelerator Laboratory, Menlo Park, California 94025, USA}
\affiliation{Department of Physics, Massachusetts Institute of Technology, Cambridge, Massachusetts 02139, USA}
\author{James R. Williams}
\affiliation{Joint Quantum Institute and Quantum Materials Center, Department of Physics, University of Maryland, College Park, MD, USA}
\affiliation{Department of Physics, Stanford University, Stanford, California 94305, USA}
\affiliation{Stanford Institute for Materials and Energy Sciences, SLAC National Accelerator Laboratory, Menlo Park, California 94025, USA}
\author{David Goldhaber-Gordon}
\email{goldhaber-gordon@stanford.edu}
\affiliation{Department of Physics, Stanford University, Stanford, California 94305, USA}
\affiliation{Stanford Institute for Materials and Energy Sciences, SLAC National Accelerator Laboratory, Menlo Park, California 94025, USA}

\begin{abstract}
A lateral Josephson junction in which the surface of a 3D topological insulator serves as the weak link should support topologically protected excitations related to Majorana fermions. 
The resulting $4\pi$-periodic current-phase relationship could be detected under high-frequency excitation by the suppression of odd Shapiro steps. 
Here, we demonstrate such devices through the self-formation of a Pd-Te superconducting layer from a telluride topological insulator, and observe suppressed first and third Shapiro steps. 
Other devices, including those where the Pd-Te layer is bolstered by an additional Al layer, show no suppression of Shapiro steps, a difference supported by simulations. 
Though we rule out the known trivial causes of suppressed Shapiro steps in our devices, we nevertheless argue that corroborating measurements and disorder-aware theoretical descriptions of these systems are needed before confidently claiming the observation of Majorana states.
\end{abstract}

\maketitle

The surface of a three-dimensional topological insulator (3D TI) hosts a non-degenerate band of massless Dirac fermions~\cite{fu2007}. Proximity to an $s$-wave superconductor (S) is predicted to mediate $p+ip$ pairing in the topological surface state, a consequence of the spin texture of the Dirac band. A Josephson junction with a TI weak link should support topologically protected gapless Andreev bound states (ABSs) known as Majorana bound states (MBSs)~\cite{fu2008}. MBSs impart a $4\pi$-periodic component to the junction's current-phase relationship, which coexists with the $2\pi$-periodic component from the spectrum of conventional ABSs at higher energies. In principle, 4$\pi$ periodicity can be detected via the fractional AC Josephson effect: junction current oscillates at half the normal Josephson frequency for a given voltage, or equivalently DC junction voltage is twice as large for a given frequency of AC current. Thus under radio frequency (RF) irradiation Shapiro steps $V=nhf/2e$ with odd $n$ should be absent.

Claims for the observation of the fractional AC Josephson effect have been made in a variety of topological systems, including nanowires with spin-orbit coupling~\cite{rokhinson2012,laroche2019}, strained 3D HgTe~\cite{wiedenmann2016}, 2D HgTe~\cite{bocquillon2017, deacon2017}, Dirac semimetals~\cite{li2018}, \ce{Bi_2Se_3}~\cite{lecalvez2019}, and \ce{(BiSb)_2Te_3}~\cite{schuffelgen2019,rosenbach2021}. 
Among these works, only in 2D HgTe has suppression of odd Shapiro steps beyond the first been observed~\cite{bocquillon2017}. Suppression of the first Shapiro step, however, can result from trivial effects in Josephson junctions, including Joule overheating~\cite{lecalvez2019} and underdamping~\cite{park2021}. Therefore the suppression of higher odd Shapiro steps is a crucial step in eliminating the ambiguity surrounding claims of $4\pi$ periodicity. Landau-Zener transitions suppress the expression of the first and higher odd Shapiro steps, but only in devices with near-unity interface transparency between superconductor and weak link~\cite{shabani2021}, offering a clear-cut way to rule out this mechanism.

\begin{figure*}[ht]
\centering
    \subfloat{\label{1a}}
    \subfloat{\label{1b}}
    \subfloat{\label{1c}}
    \subfloat{\label{1d}}
    \subfloat{\label{1e}}
    \subfloat{\label{1f}}
	\includegraphics[width=0.85\textwidth]{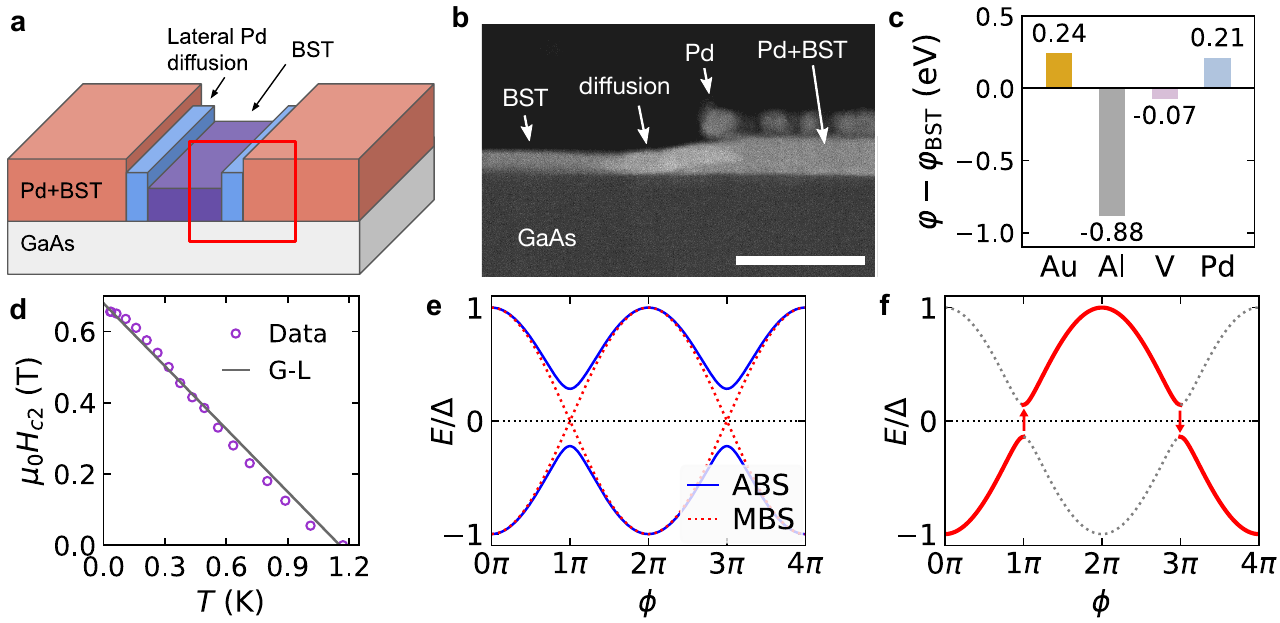}
	\caption{Junctions with a self-formed Pd-Te superconductor. (a) Schematic of a junction (not to scale). The S/TI interface is lateral, unlike most S/normal metal interfaces, where the superconductor sits on top of the normal metal. At the interface is a region where Pd diffuses laterally into the weak link. (b) Cross-sectional high-angle annular dark-field (HAADF) image of a portion of a junction indicated by the red box in (d). At left, the BST weak link. At right, the superconducting Pd-Te region, with greater film thickness due to Pd incorporated into the BST. Excess Pd has formed grains atop the film. At center, Pd has diffused laterally into the weak link by about 40~nm. Scale bar, 50~nm. (c) The work function $\varphi$ of various contact metals referenced to the work function of the BST film $\varphi_\mathrm{BST}$, measured by Kelvin probe force microscopy. (d) Out-of-plane critical field versus temperature of the Pd-Te superconductor (purple circles), and a fit (grey line) to Ginzburg-Landau theory (G-L) for out-of-plane field. (e) Theoretical spectrum of an ABS (blue line; $2\pi$ periodicity) and a MBS (red dashed line; $4\pi$-periodicity). (f) An excitation (red line) traversing a $2\pi$-periodic ABS (grey dashed line) as the phase $\phi$ across the junction evolves, with LZTs (red arrows) at $\phi=\pi$ and $3\pi$, imparting a $4\pi$-periodic component to the current-phase relationship.}
	\label{fig1}
\end{figure*}

Fabricating high-quality Josephson junctions with TI weak links is technically challenging. A number of groups have demonstrated superconducting contact to exfoliated flakes from single crystals in the \ce{Bi2Se3} family of topological insulators~\cite{sacepe2011, williams2012,veldhorst2012, kurter2015,stehno2016, ghatak2018,chen2018,kayyalha2020}. Yet even were Majorana physics confirmed, the impact of approaches reliant on individually exfoliated flakes would be limited by the need for scalability and reproducibility. Aspiring towards the scientific imperative to compile results from many devices alongside the technological goal of a scalable quantum information architecture based on Majorana modes~\cite{alicea2011}, it is vital to transition to films grown at wafer scale by molecular beam epitaxy (MBE)~\cite{shabani2016}. Yet progress in MBE-based platforms has been hampered by poor S/TI interface quality, the difficulty in protecting fragile chalcogenides during device fabrication~\cite{andersen2023}, and unwanted doping of the TI due to charge transfer across the S/TI interface.

In this work we fabricate lateral Josephson junctions with topological insulator weak links through the self-formation of a superconducting Pd-Te layer, as pioneered by Ref.~\cite{bai2020}. Using a variety of imaging techniques and low-frequency electrical measurements, we show that our fabrication process (1) yields a S/TI interface with moderate transparency, while (2) minimizing damage to the TI film and (3) nearing a work-function match between S and TI. Under radio-frequency (RF) excitation, we observe suppression of the 1st and 3rd Shapiro steps in one device, and suppression of the 1st step in more devices. Other devices express all Shapiro steps. We argue that our observations result from neither Landau-Zener transitions, junction hysteresis, nor bias-dependent resonances~\cite{mudi2021}. 
Across devices, simulations in which the $4\pi$-periodic component comprises a few percent of the total junction supercurrent generally match the observed expression of Shapiro steps.
Our results therefore constitute compelling evidence in favor of the presence of Majorana bound states in the junctions. 
Yet several differences between measurement and simulations
signal that our description of the junction dynamics is incomplete. We highlight the need for corroborating results and for further exploration of disordered S/TI interfaces ahead of definitive conclusions about Majorana states.

\subsection{Methods}

To isolate Majorana physics in TI weak-link junctions, the S/TI heterostructure must achieve sufficient electrical transparency at the S/TI interface while preserving the topological character of the TI.
The former condition, well-known from S/semiconductor structures~\cite{Takayanagi1995, xiang2006, abay2013, krogstrup2015, kjaergaard2017}, is required to open a pairing gap in the topological surface states via the proximity effect. The latter condition is specific to topological matter, and is particularly difficult due to the susceptibility of the \ce{Bi2Se3} family of materials to unwanted doping. A TI's Fermi level should lie within the bulk bandgap so that the topological surface states are not shunted by trivial bulk states. In MBE-grown films, the Fermi level is commonly tuned by adjusting the composition of ternary \ce{(Bi_xSb_{1-x})_2Te_3} (BST) and quarternary \ce{(Bi_xSb_{1-x})_2(Se_{y}Te_{1-y})_3} alloys, compensating for charged disorder including Te vacancies, Sb-Te anti-site defects, and interfacial defects~\cite{Yayu2011,Koirala2015,Salehi2019}. Yet post-growth device fabrication can introduce additional disorder, destroying the delicate charge balance or introducing mid-gap defect states~\cite{andersen2023}. Furthermore, charge transfer from the superconductor can substantially dope the TI unless their work functions are closely matched.

Two recent technologies have enabled transparent S/TI interfaces and, in turn, realization of the Josephson effect in MBE-grown TI films. One approach is to grow the TI and superconductor structures entirely in situ using stencil lithography to facilitate patterning, an approach leading to the observation of a suppressed first Shapiro step~\cite{schuffelgen2019}. In this work, we follow a second approach, the self-formation of a superconducting Pd-Te layer through the laterally-patterned ex situ deposition of Pd~\cite{bai2020}. Though in general chemical reactions with deposited metal are problematic~\cite{schuffelgen2017} (for example, depositing Al on \ce{Sb_2Te_3} might create a barrier layer of \ce{AlSb}, a $\sim 2$~eV bandgap semiconductor), here the reactivity between Pd and Te is desired.

Our devices are based on an 8 quintuple layer thick \ce{(Bi_{0.4}Sb_{0.6})2Te3} film grown by MBE on a GaAs substrate. We pattern resist masks for Josephson junctions using a low-voltage electron beam lithography process developed to impart minimal beam damage to the TI film~\cite{andersen2023, sup}. Lateral Josephson junctions are fabricated by depositing 11~nm \ce{Pd} on a resist-masked BST film in an electron beam evaporator, forming a superconducting Pd-Te alloy (with residual Bi and Sb) in exposed regions, while the BST weak link is masked (Fig.~\ref{1a}). After depositing the Pd and then liftoff, unwanted areas of BST are etched by Ar ion milling. In some devices the etched region abutts the Josephson junction weak link, terminating the transverse extent of bound states, whereas in other devices a region of BST film surrounding the weak link remains unetched; we noticed no corresponding difference in electronic transport. Our methodology and the junction geometries are described further in the \supp{}~\cite{sup}.

Junctions have geometric length of roughly $L=160$~nm (parallel to current flow; measured by scanning electron microscopy) and width $W=2$~\textmu m (transverse to current flow). Throughout this work, critical temperatures, fields, and currents are defined by the condition $R = R_N/2$ where $R_N$ is the normal state resistance. The critical temperature and field of the superconductor are determined by transport through a strip of the superconductor with no weak link. Both devices presented in the main text were fabricated simultaneously on the same chip. Three more devices on the same chip and several devices fabricated separately on other chips from the same BST film growth are discussed in the \supp{}. Separate dilution refrigerators were used for low-frequency characterization measurements of Device~1 (at Stanford Univ.; base temperature 30~mK) and Shapiro step measurements of Device~2 (at Univ. Maryland; base temperature 50~mK). All measurements used standard lock-in techniques; for Shapiro step measurements, microwave excitations were combined with low-frequency excitations at source electrodes.

\subsection{Characterization}

\begin{figure}
\centering
    \subfloat{\label{2a}}
    \subfloat{\label{2b}}
	\includegraphics[width=0.4\textwidth]{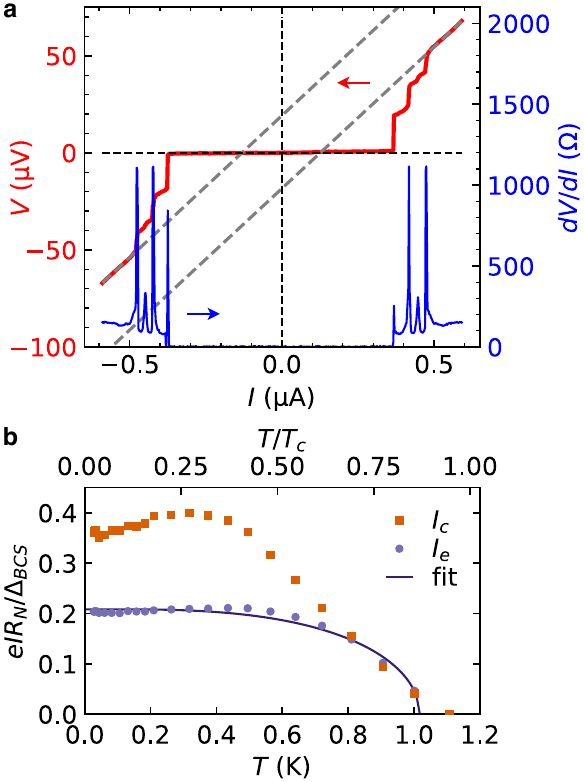}
	\caption{Junctions with Pd-Te superconducting leads under DC bias. (a) Voltage and differential resistance of Device~1 as a function of current bias. (Dashed lines) Linear fits at high bias. The excess current is determined by the intercept of these lines with $V=0$. (b) Critical current and excess current of Device~2 as a function of temperature. (Dashed line) The temperature dependence of the BCS superconducting gap fit to the excess current~\cite{sup}.}
	\label{fig2}
\end{figure}

We here summarize findings from material and device characterization measurements, which are presented in greater detail in the \supp{}.

\emph{Topological insulator}: The electronic characteristics of the BST film are determined by Hall geometry measurements at 30~mK. The film's sheet resistivity $\rho_{xx}=1.55$~k\textOmega, from $n$-type charge carriers with density $n=4.8\times 10^{12}\ \text{cm}^{-2}$, mobility $\mu=820$~cm$^2/$Vs, and elastic mean free path $\ell_e = \hbar k_F \mu/e = 15$~nm. The film has negative magnetoconductance, indicating weak anti-localization~\cite{wang2011,rosen2019}. Fitting the magnetoconductance by the Hikami-Larkin-Nagaoka formula yields a phase coherence length $\ell_\phi=850$~nm~\cite{HLN}, which provides an estimate for the inelastic mean free path $\ell_i$.

\emph{Junction characteristics}: A cross-sectional image of a portion of a junction, taken in a transmission electron microscope, is shown in Fig.~\ref{1b}. The deposited Pd diffuses vertically through the entire BST film and into the GaAs substrate. Excess Pd forms grains on top of the film. Pd diffuses laterally into the weak link by roughly 40~nm. These findings are corroborated by energy-dispersive x-ray spectroscopy, x-ray photoelectron spectroscopy, and scanning Auger electron spectroscopy.

Since Pd diffuses through the full vertical extent of the film, regions where Pd has been deposited have no remaining topological insulator layer. The direction of current flow at the edge of the weak link is therefore normal to the S/TI interface, as sketched in Fig.~\ref{1a}. This geometry differs from that of junctions based on deposited elemental superconductors, which sit atop the topological insulator so that current flows parallel to the interface plane.

\emph{Work function offsets}: Many superconductors have work functions substantially offset from that of members of the \ce{Bi_2Se_3} family. For example, the work functions of bulk \ce{Al} and \ce{Nb} are less than that of \ce{Bi_2Te_3} by nearly 1~eV~\cite{michaelson1977,takane2016}, which exceeds the 0.3~eV bulk bandgap of \ce{Bi_2Te_3}. At a transparent interface between these two materials, charge transfer should dope the TI, moving the chemical potential into the bulk conduction band and admitting topologically trivial Cooper pairing. Kelvin probe force microscopy (KPFM) confirms the work function of evaporated \ce{Al} is offset from that of BST by $-880$~meV (Fig.~\ref{1c}).

The work functions of Pd and Pd-Te alloys are substantially closer. KPFM indicates a work function offset of roughly $200$~meV between Pd-Te and BST (Fig.~\ref{1c}).

\begin{figure}
\centering
    \subfloat{\label{fig:3a}}
    \subfloat{\label{fig:3b}}
	\includegraphics[width=0.35\textwidth]{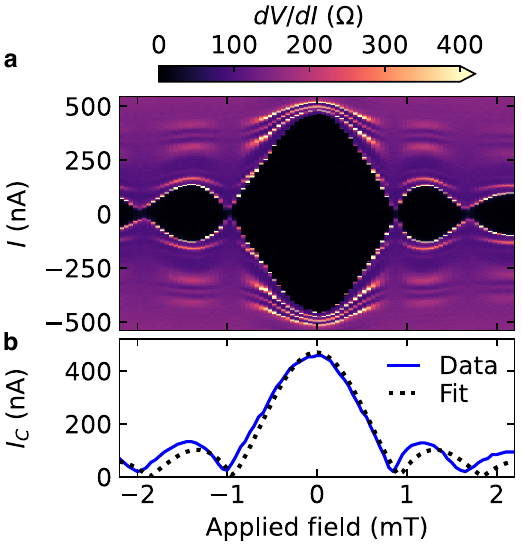}
	\caption{Device~1 in a perpendicular magnetic field. (a) Differential resistance at finite bias. (b) The extracted critical current. (Dashed line) Fit to the Fraunhofer pattern.}
	\label{fig3}
\end{figure}

\emph{Electrical properties}: The Pd-Te superconductor has critical temperature $T_c=1.17$~K, normal state sheet resistance 100~\textOmega, and critical field $\mu_0H_{c2}=655$~mT (Fig.~\ref{1d}), which implies a coherence length $\xi = 22.4$~nm through the Ginzburg-Landau relation $\xi^2 = \frac{\Phi_0}{2\pi H_{c2}}$. The devices therefore fall in the long dirty junction limit $\xi, \ell_e \ll L$.

The current-voltage relationship of Device~1 is shown in Fig~\ref{2a}. Supercurrent flows across the junction below the critical current $I_c=370$~nA. Between the critical current and roughly 500~nA are a series of sudden increases in the voltage, which are discussed in the \supp{}. At higher currents the differential resistance returns to the normal state resistance $R_N=146$~\textOmega. Extrapolating the normal section of the current-voltage relationship to zero voltage yields an excess current $I_e=136$~nA. The critical and excess currents are shown as a function of temperature in Fig.~\ref{2b}. Although Pd-Te superconductivity may not be well described by Bardeen-Cooper-Schrieffer (BCS) theory, if we take the superconducting gap as $\Delta_\text{BCS}=1.76k_BT_c$ we arrive at the dimensionless figures of merit $eI_cR_N/\Delta_\text{BCS} = 0.30$ and $eI_eR_N/\Delta_\text{BCS} = 0.10$. Naively, the latter implies a junction transparency $\tau\approx 0.25$ according to BTK theory~\cite{btk}. Device~2 has $eI_eR_N/\Delta_\text{BCS} = 0.29$, implying $\tau\approx 0.3$. We present a nuanced analysis of the junction transparency in the \supp{}. The current-voltage relationship is not hysteretic, indicating that the junction is overdamped (consistent with our estimation of the Stewart-McCumber parameter $\beta_C\sim 10^{-4}$) and that Joule overheating is not limiting the retrapping current~\cite{courtois2008}.

Fig.~\ref{fig3} shows the resistance of Device~1 in a perpendicular magnetic field. The critical current displays the typical Fraunhofer pattern, approaching zero at nonzero integer multiples of a characteristic magnetic field $B_0 = \Phi_0/L_\text{eff}W$, where $L_\text{eff}$ is the effective junction length. A fit yields $L_\text{eff}= 1.1$~\textmu m, which is significantly larger than the geometric length of the junction $L\approx 160$~nm. We find this disparity surprising, as we expect minimal flux focusing by the thin Pd-Te leads.

\begin{figure}
\centering
    \subfloat{\label{4a}}
    \subfloat{\label{4b}}
	\includegraphics[width=0.4\textwidth]{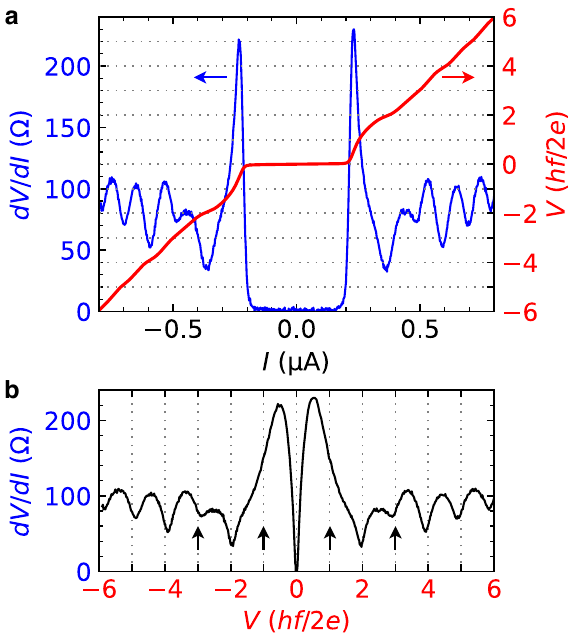}
	\caption{Current-voltage relationship in Device~2 under 4.3~GHz excitation (nominal power $-33.5$~dBm). (a) (Left) differential resistance versus current bias, and (right) DC voltage, obtained by numerical integration of the differential resistance. (b) The same data shown parametrically, emphasizing the suppression of the first and third Shapiro steps, as indicated by the arrows.}
	\label{fig4}
\end{figure}

\begin{figure*}
\centering
    \subfloat{\label{5a}}
    \subfloat{\label{5b}}
    \subfloat{\label{5c}}
    \subfloat{\label{5d}}
    \subfloat{\label{5e}}
    \subfloat{\label{5f}}
    \subfloat{\label{5g}}
    \subfloat{\label{5h}}
    \subfloat{\label{5i}}
	\includegraphics[width=\textwidth]{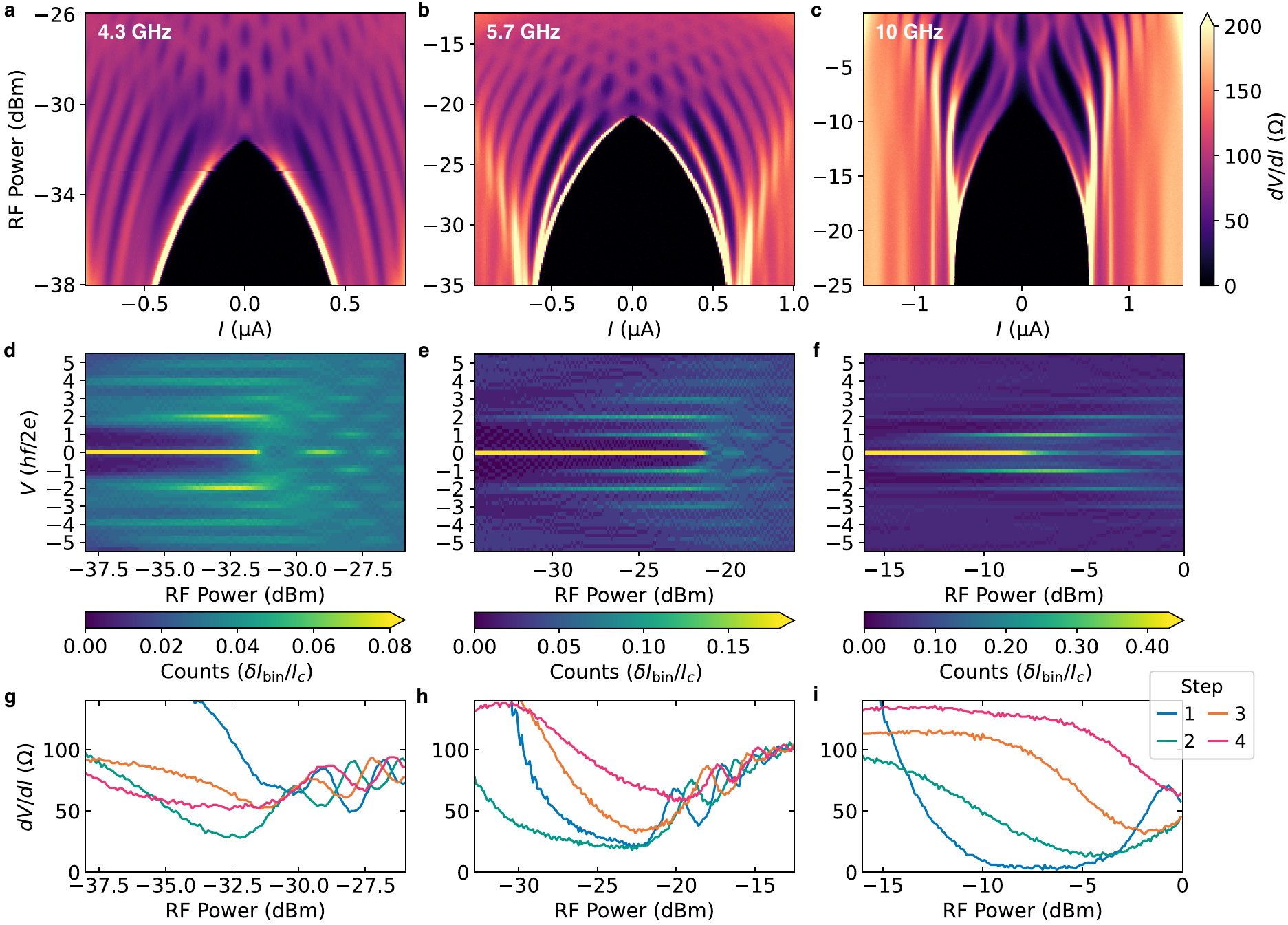}
	\caption{Shapiro steps in Device~2. (a-c) Differential resistance as a function of bias current and RF excitation power. (d-f) Corresponding histograms of measured data points within DC voltage bins, shown in normalized units $hf/2e$, with the number of counts in each bin normalized as a fraction of the critical current. The $n$th Shapiro step appears at voltage $nhf/2e$. (g-i) The value of the differential resistance at the voltage corresponding 1st through 4th Shapiro step, as a function of RF power. The frequency of RF excitation is (a, d, g) $f=$4.3~GHz, (b, e, h) 5.7~GHz, and (c, f, i) 10~GHz.}
	\label{fig5}
\end{figure*}

\subsection{Junctions under RF irradiation}

The low frequency differential resistance of Device~2 when subjected to an additional radio-frequency (RF) drive at frequency $f=4.3$~GHz is shown in Fig.~\ref{fig4}. Outside of the central zero-resistance region at low bias and low RF power are a series of regions of low differential resistance at finite DC voltage. These regions are centered at voltages $nhf/2e$ for integer $n$, expressing the $n$th Shapiro steps. The strengths of differential resistance dips associated with the 1st and 3rd Shapiro steps are suppressed relative to those of the 2nd and 4th steps.

The evolution of the differential resistance with RF excitation power is shown at different RF frequencies in Fig.~\ref{5a}-\ref{5c}. The development of Shapiro steps is clarified by the histograms Fig.~\ref{5d}-\ref{5f} formed by grouping the data points (equally spaced in DC current) into DC voltage bins, so that Shapiro steps are visible as bright streaks. At 4.3~GHz, the weights of the 1st and 3rd Shapiro steps are suppressed at low powers in comparison to those of the 2nd and 4th steps, and develop only at higher powers. The 3rd step is recovered as the RF frequency is increased to 5.7~GHz, as is the 1st step at 10~GHz. 
Fig.~\ref{5g}-\ref{5i} present another visualization of the formation of Shapiro steps by showing the differential resistance at the voltages corresponding to the first four Shapiro steps as a function of RF power. 
At 4.3~GHz, even steps form at lower RF power than odd steps, as indicated by the lower differential resistance, whereas at 10~GHz, the steps form sequentially. 
We present additional visualizations of these data in Section~S8 of the \supp{}, along with data from three further devices, in two of which we observe suppression of the first Shapiro step.

The suppression of odd Shapiro steps is an expected signature of the presence of a MBS. The ABS spectrum of a conventional Josephson junction is $2\pi$-periodic, leading to a $2\pi$-periodic current phase relationship and Shapiro steps at voltages $V=nhf/2e$ for integer $n$. The $2\pi$-periodicity results from an avoided crossing of the two branches of the lowest-energy ABS at $\phi=\pi$ (Fig.~\ref{1e}), a consequence of scattering in the junction or at interfaces. However, contact to a superconductor is expected to induce effective $p$-wave pairing in a topological insulator~\cite{fu2008}, and a junction between $p$-wave superconductors should support a pair of Majorana bound states~\cite{kitaev2001,fu2009,lutchyn2010}. At $\phi=\pi$ and $3\pi$, the two Majorana bound states do not couple and are therefore bound to zero energy. Because these states differ in fermion parity (unlike ABS, which tunnel charge $2e$), this picture should be valid provided the Majorana modes are spatially separated and there are no quasiparticle excitations~\cite{kane2015}, requiring $k_BT\ll \Delta_\text{BCS}$.

The junction's current-phase relationship therefore gains a $4\pi$-periodic component, which coexists with the $2\pi$-periodic term~\cite{kwon2004}. At high RF power and frequency, the $2\pi$ component is expected to dominate the junction dynamics, but at low RF power and frequency, the $4\pi$ component may dominate, leading to a suppression of Shapiro steps with odd $n$~\cite{lecalvez2019}. 

The observation of suppressed odd Shapiro steps does not necessarily imply the presence of a Majorana mode, yet the known alternative mechanisms do not apply to our devices. First, the lowest Shapiro steps can be suppressed at low RF power in underdamped junctions~\cite{park2021}, as well as in overdamped junctions with substantial self-heating~\cite{dececco2016,shelly2020}. The small capacitance and lack of hysteresis in our junctions is not consistent with these effects; furthermore, these effects cannot suppress the third step while leaving the second step intact.

Second, suppression of the first and third Shapiro steps due to Landau-Zener transitions (LZTs) between upper and lower ABS branches near $\phi=\pi$ (Fig.~\ref{1f}) has been predicted~\cite{Billangeon2007, dominguez2012} and recently observed in a lateral Josephson junction based on a topologically trivial \ce{InAs} weak link~\cite{shabani2021}, enabled by exceptionally high interface transparency. The probability of a LZT is
\begin{equation}
    P = \exp\left(-\frac{\pi (1-\tau)\Delta}{eV}\right),
\end{equation}
which is significant only when $\tau\approx1$~\cite{averin1995}. Our junctions have $\tau$ of roughly~$0.3$, far from the ultra-high-transparency regime in which LZTs occur.

Third, recent computational work suggests that the $n$th Shapiro step can be suppressed by unwanted device resonances if the resonances occur at DC voltages at or near $nhf/2e$~\cite{mudi2021}. We do observe resonances above the critical current of our devices. The resonances appear at fixed voltages with varying field and temperature, even close to $T_c$, which is inconsistent with multiple Andreev reflection and Fiske resonance (see Section~S6 of the \supp{}). Instead, we suggest that the resonances arise from radiative coupling between the device and cavity modes of the cryostat or of the electrical wiring. Yet, regardless of their origin, resonances do not clearly explain the suppression of odd Shapiro steps in our devices, as we observe suppression of the first step throughout the range 2.5~GHz (the lowest frequency at which Shapiro steps clearly develop) to 7~GHz, which is inconsistent with resonances appearing at fixed voltages. Therefore, we exclude hysteresis, LZTs, and resonances as alternatives to the presence of MBS to explain the observation of suppressed Shapiro steps.

In Section~S9 of the \supp{}, we present numerical simulations of the Device's resistance under microwave irradiation based on the resistively shunted junction (RSJ) model
\begin{equation}\label{eq:RSJ}
\frac{V}{R_N} + I_{2\pi}\sin(\phi) + I_{4\pi}\sin(\frac{\phi}{2}) = I_\mathrm{DC} + I_\mathrm{RF}\sin(ft),
\end{equation}
where $V=\hbar \dot \phi/2e$ is the voltage across the junction, $\phi$ is the phase across the junction, $R_N$ is the normal-state resistance, $I_{2\pi\, (4\pi)}$ is the $2\pi$ ($4\pi$)-periodic portion of the critical current, $f$ is the microwave irradiation frequency, and $I_{\mathrm{DC}\, (\mathrm{RF})}$ are the DC and microwave-frequency current biases, respectively~\cite{snyder2018, mudi2021}. Values for $R_N$ and the total critical current $I_C=I_{2\pi}+I_{4\pi}$ are taken from junction characterization measurements. Simulations of Device~2 feature noticeable suppression of odd Shapiro steps when $I_{4\pi}$ is just a few percent of $I_C$, confirming the plausibility of interpreting suppressed Shapiro steps as a signature of MBS. Experimental and simulated results differ in the widths of the regions between adjacent steps. The differences are discussed in Section~S9 of the \supp{} and may indicate that the current-phase relationship in Eq.~\ref{eq:RSJ} is an incomplete model of the junction dynamics.

\subsection{Junctions with Al}

A drawback of our methodology is the short coherence length of the Pd-Te superconductor, which places our devices near the long junction regime~\cite{sup}. We attempt to bolster the superconducting coherence length of the junction leads by evaporating 65~nm Al immediately after depositing 5~nm Pd on the BST film. The hybrid Al/Pd-Te superconductor has critical temperature 780~mK (whereas evaporated Al alone has $T_c$ of roughly 1.3~K), and critical field 13~mT. Compared to junctions with Pd-Te alone, junctions with hybrid superconducting leads have similar low-frequency transport characteristics, but lower critical currents and excess currents (in some devices, $I_e<0$)~\cite{sup}.

Since the coherence length is increased in devices with the hybrid superconductor, but the junction length is unchanged, the junctions should support fewer trivial ABS~\cite{bagwell1992} coexisting with the single MBS. We would therefore expect a proportionally larger $4\pi$-periodic component, leading to more prominent suppression of odd Shapiro steps. Instead, we observe expression of all Shapiro steps in all measured devices (data are shown in Section~S8 of the \supp{}). 

In simulations of the devices with Al, presented in Section~S9 of the \supp{}, the $4\pi$-periodic bound states must constitute a higher portion of the total critical current---tens of percent---for the suppression of odd Shapiro steps to be visible as compared to the devices with Pd-Te alone. This difference is due to the lower $I_C R_N$ product for devices with Al~\cite{dominguez2012, sup}. The simulations therefore indicate that MBS could possibly be present in all devices, yet are revealed by Shapiro measurements only in devices with the highest transparencies.

In some devices, Shapiro steps are also expressed at fractional multiples of $hf/2e$~\cite{sup}. Shapiro steps at multiples of $hf/4e$ have been observed in junctions with exceptionally high transparency~\cite{ueda2020}, reflecting the skewed current-phase relationship in this regime~\cite{likharev}, yet our devices have substantially lower transparency. Further work is needed to understand the origin of the fractional steps in our devices.

\subsection{Conclusion}

In this work, we fabricate Josephson junctions with topological insulator weak links using a self-formed superconductor to provide good interface transparency while minimizing damage to the TI. We observe suppression of the first and third Shapiro steps under low power and low frequency RF excitation, a hallmark of the fractional AC Josephson effect consistent with the existence of Majorana bound states: topologically protected gapless Andreev bound states. To the best of our knowledge, our data are inconsistent with topologically trivial sources for the suppression of Shapiro steps and present the strongest evidence to date of $4\pi$~periodicity in a 3D TI-based Josephson structure.

Yet difficulty of obtaining results that are reproducible between devices and that precisely match simulation raises questions for such an interpretation. More work is needed to understand how the structure of the S/TI interface influences Andreev spectra and to confirm that observations of suppressed Shapiro steps reflect $4\pi$ periodicity rather than its mere mimicry. Provision of full data sets without post-selection, as we have done here, is needed for the community to reach reliable conclusions regarding the existence of Majorana modes.

Measurements of the fractional AC Josephson effect, such as those presented here, can be complemented by corroborative probes of the presence of Majorana modes. Majorana bound states should exist in the core of artificial vortices in proximitized TIs, namely regions of TI surrounded by a superconductor wound by a phase $2\pi$. An artificial vortex may be realized as a flux-biased superconducting ring or as a phase-biased superconducting trijunction~\cite{fu2008}. In both geometries, the presence of a Majorana bound state could be corroborated by a zero-bias peak in the conductance spectroscopy of a tunnel probe, analogous to the features sought in semiconducting nanowires~\cite{mourik2012} and vortex cores of candidate intrinsic $p$-wave superconductors~\cite{wang2018,zang2021}. Yet the necessary tunnel contacts to \ce{Bi_2Se_3}-class topological insulators are difficult to fabricate {\em ex situ}. In the \supp{}, we propose a device based on self-formed superconducting contacts that could address this challenge.

Finally, the Pd-BST heterostructure developed in this work could be extended to realize transparent superconducting contact to the quantum anomalous Hall system. Here, the BST would be replaced with its ferromagnetic analogue, the quantum anomalous Hall insulator \ce{Cr-(BiSb)_2Te_3}, and the Pd-Te superconductor would be bolstered by deposition of a large-gap superconductor like \ce{Nb} immediately following deposition of the \ce{Pd} (likely needed to overcome depairing due to magnetic exchange).

{\em Data availability}: The full dataset is provisioned along with analysis code at \url{https://zenodo.org/doi/10.5281/zenodo.5893506}.

L. T., P. Z., and K. L. W. developed and grew the BST film. I. T. R. and M. P. A. fabricated and characterized the devices and conducted low-frequency measurements. C. J. T. and J. R. W. measured the devices under RF irradiation. Y. Li, Y. Liu, and Y. C. conducted TEM measurements. I. T. R., C. J. T., M. P. A., E. M., M. A. K., J. R. W., and D. G.-G. analyzed the data. I. T. R. wrote the manuscript with contributions from all authors.

I. T. R., M. P. A., E.M., and Y. Li were supported by the U.S. Department of Energy, Office of Science, Basic Energy Sciences, Materials Sciences and Engineering Division, under Contract DE-AC02-76SF00515. I. T. R. additionally acknowledges support from the ARCS foundation. P. Z., L. T., and K. L. W. were supported by the U.S. Army Research Office MURI program under Grant No. W911NF-16-1-0472. AC Josephson measurements of the devices were sponsored by the grant Army Research Office Award W911NF-18-2-0075. Infrastructure and cryostat support were funded in part by the Gordon and Betty Moore Foundation through Grant No. GBMF3429. Part of this work was performed at the nano@Stanford labs, which are supported by the National Science Foundation as part of the National Nanotechnology Coordinated Infrastructure under award ECCS-1542152. 

While preparing this manuscript, we became aware of a work with similar methodology~\cite{bai2021novel}.

\nocite{castroneto2009,Zhang2009,Yayu2011,dubos2001,bai2020,ko2011VO2,michaelson1977,furusaki1999,aminov1996,chrestin1997,likharev,bagwell1992,btk,otbk,park2021,andersen2023,wang2018,zang2021,mourik2012,brahlek2016,jiang2018,Salehi2019,ren2019}
\bibliography{references.bib}

\end{document}


\title{Supplemental Material for ``Fractional AC Josephson Effect in a Topological Insulator Proximitized by a Self-Formed Superconductor''}
\author{Ilan T. Rosen}
\affiliation{Department of Applied Physics, Stanford University, Stanford, California 94305, USA}
\affiliation{Stanford Institute for Materials and Energy Sciences, SLAC National Accelerator Laboratory, Menlo Park, California 94025, USA}
\author{Christie J. Trimble}
\affiliation{Joint Quantum Institute and Quantum Materials Center, Department of Physics, University of Maryland, College Park, MD, USA}
\author{Molly P. Andersen}
\affiliation{Department of Materials Science and Engineering, Stanford University, Stanford, California 94305, USA}
\affiliation{Stanford Institute for Materials and Energy Sciences, SLAC National Accelerator Laboratory, Menlo Park, California 94025, USA}
\author{Evgeny Mikheev}
\affiliation{Department of Physics, Stanford University, Stanford, California 94305, USA}
\affiliation{Stanford Institute for Materials and Energy Sciences, SLAC National Accelerator Laboratory, Menlo Park, California 94025, USA}
\author{Yanbin Li}
\affiliation{Department of Materials Science and Engineering, Stanford University, Stanford, California 94305, USA}
\author{Yunzhi Liu}
\affiliation{Department of Materials Science and Engineering, Stanford University, Stanford, California 94305, USA}
\author{Lixuan Tai}
\affiliation{Department of Electrical Engineering, University of California, Los Angeles, California 90095, USA}
\author{Peng Zhang}
\affiliation{Department of Electrical Engineering, University of California, Los Angeles, California 90095, USA}
\author{Kang L. Wang}
\affiliation{Department of Electrical Engineering, University of California, Los Angeles, California 90095, USA}
\author{Yi Cui}
\affiliation{Department of Materials Science and Engineering, Stanford University, Stanford, California 94305, USA}
\affiliation{Stanford Institute for Materials and Energy Sciences, SLAC National Accelerator Laboratory, Menlo Park, California 94025, USA}
\author{M. A. Kastner}
\affiliation{Department of Physics, Stanford University, Stanford, California 94305, USA}
\affiliation{Stanford Institute for Materials and Energy Sciences, SLAC National Accelerator Laboratory, Menlo Park, California 94025, USA}
\affiliation{Department of Physics, Massachusetts Institute of Technology, Cambridge, Massachusetts 02139, USA}
\author{James R. Williams}
\affiliation{Joint Quantum Institute and Quantum Materials Center, Department of Physics, University of Maryland, College Park, MD, USA}
\affiliation{Department of Physics, Stanford University, Stanford, California 94305, USA}
\affiliation{Stanford Institute for Materials and Energy Sciences, SLAC National Accelerator Laboratory, Menlo Park, California 94025, USA}
\author{David Goldhaber-Gordon}
\affiliation{Department of Physics, Stanford University, Stanford, California 94305, USA}
\affiliation{Stanford Institute for Materials and Energy Sciences, SLAC National Accelerator Laboratory, Menlo Park, California 94025, USA}
\email{goldhaber-gordon@stanford.edu}

\maketitle

\renewcommand{\thetable}{S\arabic{table}}
\renewcommand{\thefigure}{S\arabic{figure}}
\renewcommand{\thesection}{S\arabic{section}}
\renewcommand{\thesubsection}{\arabic{subsection}}
\renewcommand{\theequation}{S\arabic{equation}}
\renewcommand{\thepage}{S\arabic{page}}

\setcounter{secnumdepth}{2}
\setcounter{equation}{0}
\setcounter{figure}{0}
\setcounter{section}{0}
\setcounter{page}{1}

\renewcommand{\citenumfont}[1]{S#1}
\renewcommand{\bibnumfmt}[1]{[S#1]}

\tableofcontents

\clearpage

\section{Methodology}

\emph{Film growth}: The topological insulator thin film used in this work was a high-quality single crystalline \ce{(Bi_{0.4}Sb_{0.6})2Te3} film 8 quintuple layers (QLs) in thickness. The film was grown on a semi-insulating GaAs (111)B substrate in a Perkin-Elmer ultra-high-vacuum molecular beam epitaxy system. The substrate was annealed at 580\degree~C in a Te-rich environment to remove the native oxide. The film was grown with the substrate held at 200\degree~C with the Bi, Sb, and Te source shutters simultaneously open. Growth was monitored using in situ reflection high-energy electron diffraction (RHEED).

\emph{Fabrication}: Devices were fabricated using a lift-off process developed to minimize damage to the TI film while still allowing high-resolution patterning. The Josephson junctions were patterned using electron beam lithography. A3 PMMA 950 resist was spin coated and baked at 80\degree~C for 300~s, chosen to avoid thermal damage. The resist was patterned by electron beam lithography at a 10~kV accelerating voltage and a 100~$\mu$C/cm$^2$ dose. Resist was developed in 1:3 methyl isobutyl keytone:isopropanol. Junction lead metal was deposited in an electron beam evaporator following a brief \emph{in situ} Ar ion mill cleaning step. For devices with Pd-Te alone, 11~nm Pd was deposited. For devices with an additional Al layer, 5~nm Pd and 65~nm Al was deposited.

Following deposition of the junction leads and lift-off, the TI film was etched and bondpads were deposited. These steps used maskless photolithographic patterning. For each patterning step, a hexamethyldisilazane adhesion layer was spin coated, followed by Megaposit SPR 3612 photoresist and a pre-exposure bake of 80\degree~C for 300~s. Photoresist was exposed using a 385~nm direct write at 120~mJ/cm$^2$ and was developed in Microposit developer CD-30 for 35~s. The device mesas were defined after patterning by etching the surrounding film with an Ar ion mill. Pads for wirebonding were made after patterning by a brief in situ Ar ion mill cleaning step, and then evaporating 5~nm Ti and 90~nm Au, followed by liftoff. \ce{Al} bondpads were used in devices having \ce{Al} atop the \ce{Pd} leads to avoid \ce{Au-Al} intermetallics. In some cases, devices were coated with a protective dielectric layer to reduce aging. Alumina was deposited using 400 cycles of thermal atomic layer deposition (ALD) at 60\degree~C. After ALD, the bondpads were exposed by etching the alumina with Microposit developer CD-26 (tetramethylammonium hydroxide based, metal ion free) after patterning. Metal was evaporated using a Kurt Lesker electron beam evaporator with an in situ Ar ion source. ALD used trimethylaluminum precursor and water as the oxidizer in a nitrogen purged vacuum chamber.

\emph{DC measurements}: Devices were studied with quasi-four terminal electrical measurements using a typical lock-in measurement setup in a dilution refrigerator at a base temperature of 30~mK. Devices were current biased with typically a 1~nA AC bias, applied to the device by sourcing 1~V RMS across a 1~G$\Omega$ resistor, which was added to a DC current bias applied by sourcing voltage across a 10~M$\Omega$ resistor. The current traveling through the device and out of the drain terminal was amplified with an Ithaco 1211 current preamplifier with 20~$\Omega$ input impedance, set to $-10^6$~V/A gain. Voltage was amplified using NF Corporation LI-75A differential voltage preamplifiers with $10^2$ gain. The AC current and voltage were measured using Stanford Research Systems SR830 digital lock-in amplifiers, and the DC current and voltage were measured using Agilent 34401A digital multimeters. The AC excitation frequency was $\sim 10$~Hz; the low frequency was necessitated by the RC constant of low-temperature electrical filtering (copper powder filter plus multi-pole low-pass RC filters).

\emph{RF excitation}: For measurements with RF excitation, devices were studied with quasi-four terminal electrical measurements in a dilution refrigerator at a base temperature of roughly 50~mK. Devices were current biased with typically a 10~nA low-frequency AC bias, which was added to a DC current bias. RF excitation up to 10~GHz was supplied to an electrical lead on the devices using a synthesizer through a bias-tee located on the chip carrier. AC voltage was measured using a Stanford Research Systems SR860 digital lock-in amplifier. DC voltage was determined through numerical integration of the AC voltage. The RF power presented here is the nominal output power of the synthesizer and is not corrected for attenuation and losses.

\emph{Cross-sectional electron microscopy}: For cross-sectional imaging, a test device was fabricated similarly to the other devices presented in the main text. The cross-section TEM specimen was prepared using a FEI Helios NanoLab 600i DualBeam Focused Ion Beam/Scanning Electron Microscopy (FIB/SEM). A carbon protection layer was deposited using an electron beam before the deposition of a Pt protection layer with a Ga+ ion beam. The lamella was then lifted out and milled using the ion beam at energy 30~keV. For final cleaning of the specimen, the energy of the ion beam was gradually reduced to 5~keV and eventually 2~keV. TEM characterization used a FEI Titan environmental (scanning) transmission electron microscope (E(S)TEM) operated at 300~kV. The microscope was equipped with an aberration corrector in the image-forming (objective) lens, which was tuned before each sample analysis. 

\emph{Sequence of measurements}: The film was grown at UCLA and was shipped to Stanford, where it was stored prior to device fabrication in a vacuum desiccator covered with PMMA for protection. Devices were fabricated at Stanford University, and were characterized at low frequency shortly thereafter. The data shown in Figs.~2(a) and 3 were taken at this point. The chip was then sent to the University of Maryland for high-frequency measurements, including the data shown in Figs.~4 and 5. The chip was then returned to Stanford University for additional low-frequency characterization, at which point the data shown in Fig.~2(b) was taken. SEM imaging of the devices followed all other measurements.

The Josephson junction Device~1 was found to be shorted at the University of Maryland, which is the reason we present no Shapiro step measurements in this device. Devices~3-5 were found to be shorted when the chip was returned to Stanford University, which is the reason we do not present additional characterization data for these junctions. We are uncertain what caused the junctions to become shorted; a likely possibility is gradual lateral diffusion of Pd into the weak link, perhaps accelerated by the conditions experienced during shipment.

\clearpage

\section{Imaging and characterization}

The effective electron mass in the BST film is $m^* = \hbar k_F/v_F = 0.078m_e$, where $m_e$ is the mass of an electron, $k_F = \sqrt{\pi n/2}$ (the factor of two is to account for top/bottom surface degeneracy), and assuming the Fermi velocity $v_F=4\times 10^5\ \text{m}/\text{s}$~\cite{castroneto2009,Zhang2009,Yayu2011}. The elastic mean free path is $\ell_\text{MFP}^e = \mu m^* v_F/e =\mu \hbar \sqrt{\pi n/2}/e = 15$~nm, the parameters of which are determined by Hall measurements.

The Thouless energy for a junction with $L=150$~nm is $E_{\rm Th}=\hbar D / L^2=\SI{44}{\micro eV}$ at $T=30$~mK, where the diffusion constant is $D=v_F \ell_\text{MFP}^e/4$.
The thermal length is $L_T = \sqrt{\hbar D/2\pi k_B T}=210$~nm. Since $L$ and $L_T$ are comparable, we expect the critical current to be a weak function of the junction length.
As $e\Delta/E_{\rm Th}\approx 4$, and $L\gg \ell_\text{MFP}^e$, we categorize our junctions as diffusive junctions that approach but are not fully in the long-junction regime. Here, both energy scales $e\Delta$ and $E_{\rm Th}$ impact the critical currents and quasiparticle spectra of the junctions~\cite{dubos2001}.

The critical temperature of the self-formed Pd-Te superconductor is highly dependent on \ce{Pd} deposition parameters. In the main text, we presented devices formed with evaporated \ce{Pd}. The superconductor in these devices has critical temperature $T_c=1.17$~K. In two other chips fabricated with similar deposition parameters we found $T_c=0.7$~K and 1.2~K. 

The devices studied in Ref.~\cite{bai2020} used sputtered, rather than evaporated, \ce{Pd}, and had $T_c = 0.67-1.2$~K. Devices we fabricated with sputtered \ce{Pd} had slightly lower critical temperatures $T_c = 0.4-0.7$~K.

The upper critical field of a 2D superconductor under in-plane and out-of-plane field in Ginzburg-Landau theory by
\begin{equation}
    H_{c2}^\parallel = \frac{\Phi_0 \sqrt{12}}{2\pi \xi d}\sqrt{ 1- \frac{T}{T_c}},
\end{equation}
\begin{equation}
    H_{c2}^\perp = \frac{\Phi_0}{2\pi \xi^2}\left( 1- \frac{T}{T_c}\right),
\end{equation}
where $\Phi_0=h/2e$ is the superconducting flux quantum, $\xi$ is the Ginzburg-Landau coherence length at zero temperature and $d$ is the thickness of the film. 

\begin{figure}[h!]
    \centering
    \includegraphics[width=\textwidth]{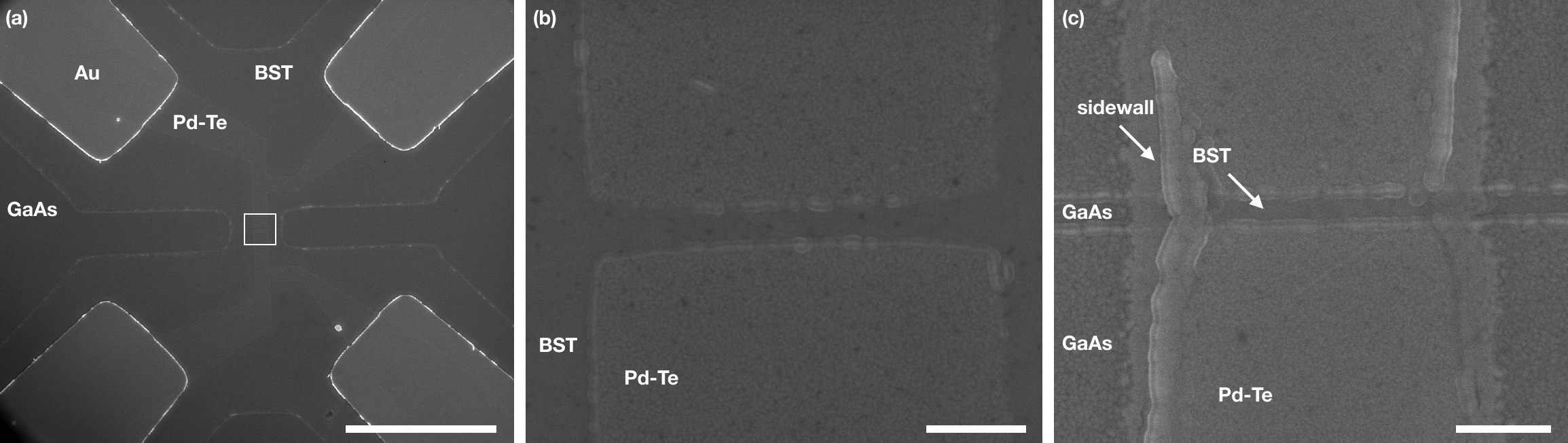}
    \caption{SEM images of junctions. (a) Wide image of Device~2. Four Au leads connect the Pd-Te superconducting leads to wirebond pads for four-terminal measurements. The Josephson junction is at center. Extraneous regions of BST are etched, revealing the GaAs substrate underneath. Scale bar, 20~\textmu m. (b) Image of the Josephson junction of Device~2. The image region is indicated by the box in (a). Scale bar, 500~nm. (c) Image of the Josephson junction of Device~1, demonstrating the etched geometry. Sidewalls from the etch abut the transverse extent of the junction. In the etched region, shadows are visible from the preceding Pd deposition. Scale bar, 500~nm.}
    \label{sfig:SEM}
\end{figure}

\begin{figure}
    \centering
    \includegraphics[width=0.7\textwidth]{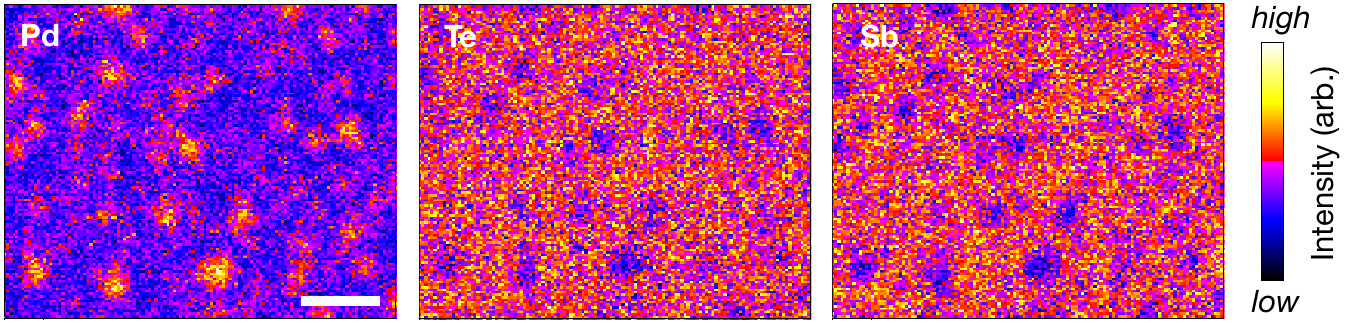}
    \caption{Image of Pd, Te, and Sb Auger emission peaks in a region where Pd has been deposited on BST. Scale bar, 200 nm.}
    \label{sfig:AESlead}
\end{figure}

\begin{figure}
    \centering
    \includegraphics[width=0.9\textwidth]{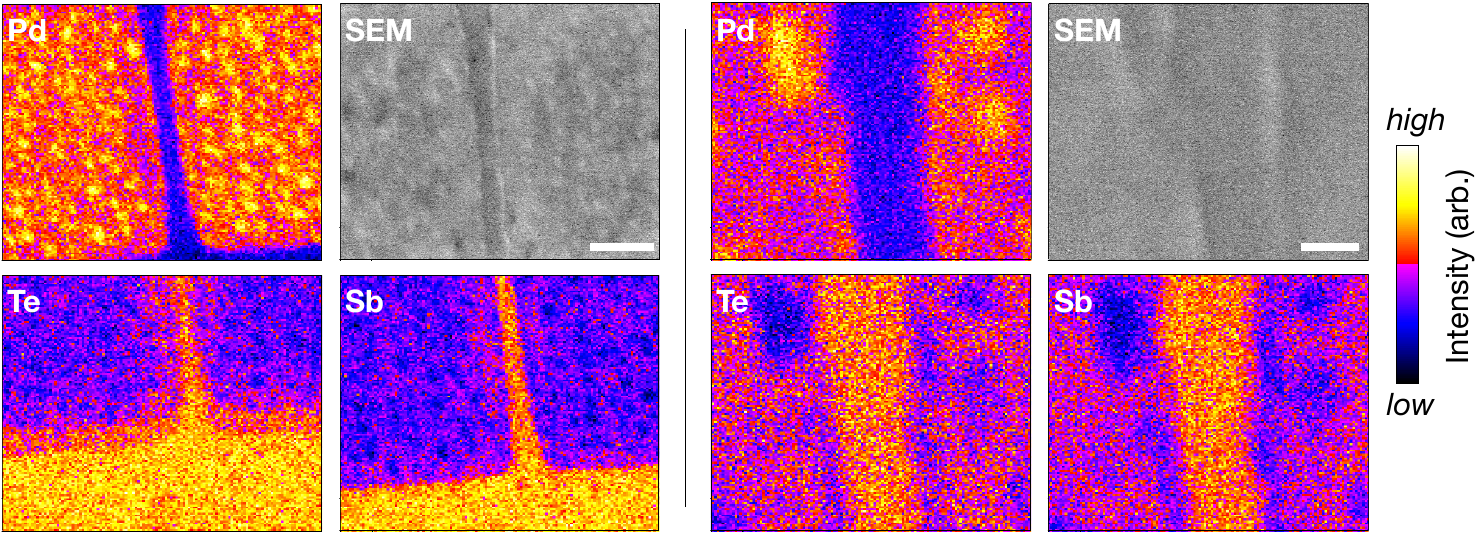}
    \caption{Image of Pd, Te, and Sb Auger emission peaks in a junction geometry. The weak link (vertical band at center) appears as a Pd-deficient, Te-rich region. Scanning electron microscopy (SEM) images provide topographical context. Left, wide view (scale bar, 400 nm). Right, detail (scale bar, 100 nm). Offsets between images are due to drift.}
    \label{sfig:AESjunc}
\end{figure}

Scanning electron microscopy (SEM), transmission electron microscopy (TEM), and accompanying energy dispersive spectroscopy (EDS) images are presented in Figs.~\ref{sfig:SEM}, \ref{sfig:TEM}, and \ref{sfig:EDS}, respectively. Auger electron spectroscopy (AES) of a region where Pd has been deposited on BST (Fig.~\ref{sfig:AESlead}) and of a weak link between two Pd-Te leads (Fig.~\ref{sfig:AESjunc}) reveals that regions of BST exposed to BST contain Pd throughout, but excess Pd forms grains on the film. The weak link lacks Pd, while the Pd-Te leads appear Te-deficient compared to the weak link because the Pd dominates the Auger emission, which has few-nm surface selectivity. Pd (3~nm) was deposited by sputtering.

AES maps were acquired on a PHI Auger 700. Palladium data was taken via the Pd MNN peak at 327.5~eV, antimony via the Sb MNN peak at 453.9~eV, and tellurium via the Te MNN peak at 483.9 eV. Data in Fig.~\ref{sfig:AESlead} and Fig.~\ref{sfig:AESjunc} (right) was taken at 25~kV with a 10~nA current. Data in Fig.~\ref{sfig:AESjunc} (left) was taken at 10~kV with a 10~nA current. All data analysis was performed with MultiPak, using the S9D9 smoothing and differentiating routines where applicable. 

\begin{figure}
    \centering
    \includegraphics[width=0.9\textwidth]{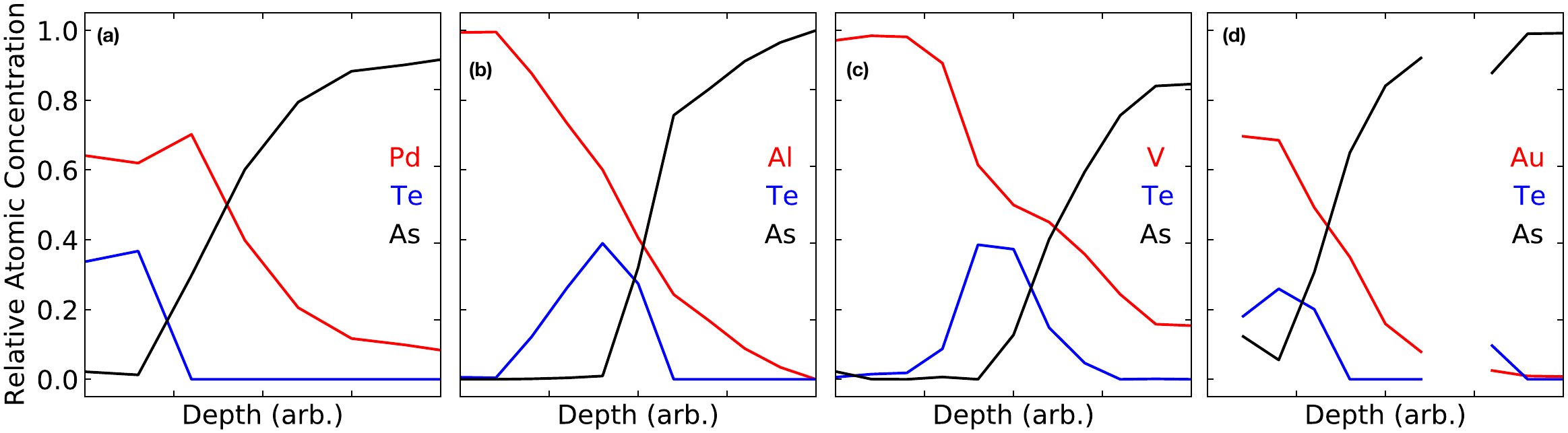}
    \caption{Relative concentration of several metals to Te, and As as a function of depth, measured by x-ray photoelectron spectroscopy. The depth profile is achieved by measuring after etching the surface by Ar ion sputtering. The etch rate is not calibrated, but the crossover from Te to As indicates that from the BST film to the GaAs substrate. (a) 4~nm Pd on BST. (b) 8~nm Al on BST. (c) 6~nm V on BST. (d) 6~nm Au on BST.}
    \label{sfig:XPS}
\end{figure}

\begin{figure*}[h!]
\centering
	\includegraphics[width=0.65\textwidth]{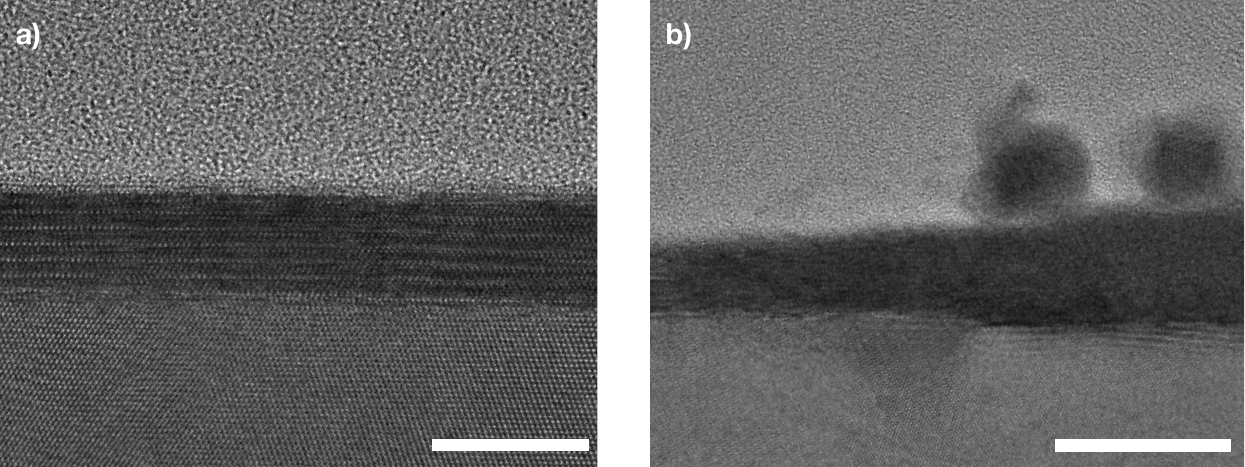}
	\caption{Cross-sectional TEM imaging. (a) The BST film (dark band at center), atop the GaAs substrate (at bottom). Scale bar, 10~nm. (b) Like (a), but Pd has been deposited atop the rightmost portion of the BST film. Scale bar, 20~nm. The sample was coated with Cu for protection prior to cross-sectioning (visible at top).}
	\label{sfig:TEM}
\end{figure*}

\begin{figure}
\centering
	\includegraphics[width=0.25\textwidth]{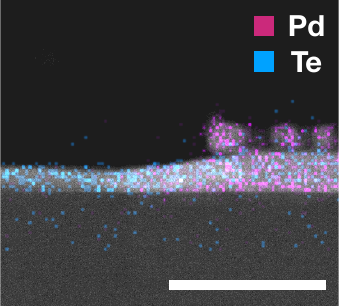}
	\caption{Cross-sectional energy dispersive spectroscopy image overlaid on the HAADF image from Fig.~1(b) of the main text. Te (blue) is found throughout the film. Pd (pink) diffuses throughout the film's full thickness, and laterally into the weak link by about 40~nm. Scale bar, 50~nm.}
	\label{sfig:EDS}
\end{figure}

X-ray photoelectron spectroscopy (XPS; Fig.~\ref{sfig:XPS}) on a BST film with 4 nm electron-beam evaporated Pd shows that the Pd diffuses throughout the entire BST film and into the GaAs substrate. The maximum concentration of Pd occurs after the maximum concentration of Te; this indicates the Pd penetrates throughout the BST film and even into the GaAs substrate. Comparable data is also shown for BST with thin electron-beam evaporated layers of Al, V, and Au on top. The maximum concentration of both Al and V occurs well before the maximum concentration of Te, indicating that neither metal penetrates through the BST film. Instead, the Al and V layers remain on top of the BST layer. Interestingly, the Au depth profile suggests that gold also penetrates into the BST film. 

Depth profiles were created by timed sputtering with Ar ions. While sputtering can affect the apparent depth of atomic species in a layered structure by resputtering and/or forcing some atoms deeper than others, the XPS depth profile for Pd matches well with the cross-sectional STEM-EDS depth profile (Fig.~\ref{sfig:EDS}) which did not involve top-down sputtering at all. We therefore conclude that the act of sputtering did not itself substantially affect the XPS depth profile data. 

XPS data was taken on a PHI, Inc. VersaProbe3 with Al-K$\alpha$ anode. Peaks were acquired with a 224~eV pass energy, 0.8~eV step size, and 200x200~\textmu m$^{2}$ spot size. Ar ion sputtering was performed in 60~s (Al, V, Au) or 90~s (Pd) increments at 500~V with a 0.6~\textmu A current and 2x2~mm raster size. Analysis was performed with MultiPak after charge compensation via shifting the C1s peak to 284.8~eV.Tellurium data was extracted from the 3d3/2 and 3d5/2 peaks ($\sim$575 eV), arsenic from the 2p3 peak ($\sim$1325 eV), palladium from the 2d3/2 and 3d5/2 peaks ($\sim$340 eV), aluminum from the 2p peaks ($\sim$75 eV), vanadium from the 2p3 peaks ($\sim$517 eV), and gold from the 4f5/2 and 4f7/2 peaks ($\sim$91 eV).

\begin{figure*}[h!]
\centering
	\includegraphics[width=0.65\textwidth]{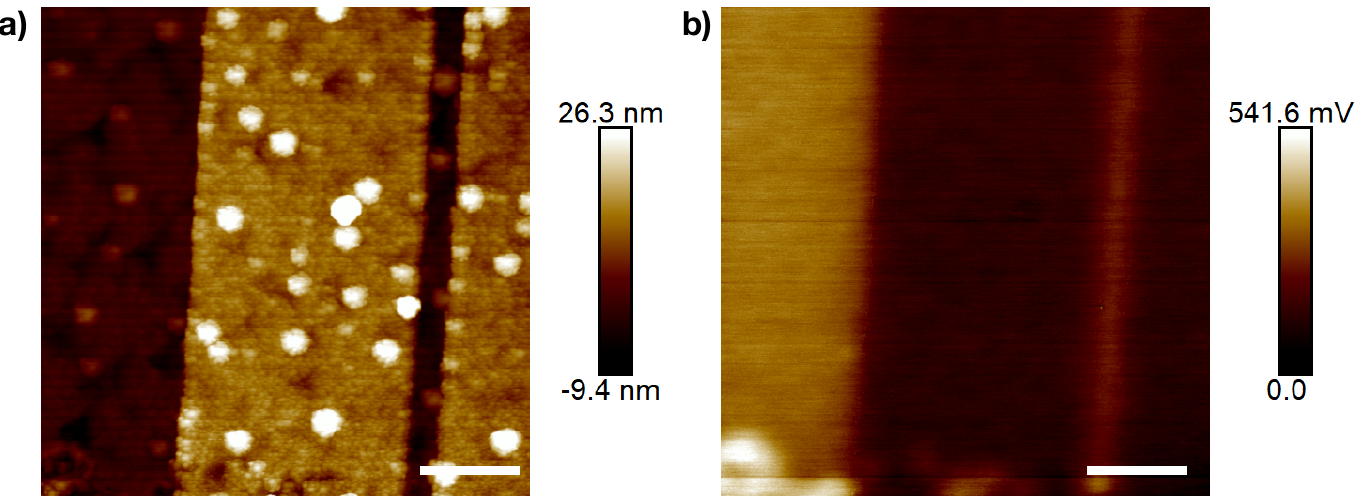}
	\caption{KPFM image of a test junction structure formed by evaporation of 11~nm Pd on BST. (a) Topography. (b) Work function measurement. Pd has been deposited in the regions at center and at far right; the region at far left and the narrow strip between the Pd regions is bare BST. Scale bar, 500~nm.}
	\label{sfig:kpfm_img}
\end{figure*}

\begin{figure*}[h!]
\centering
	\includegraphics[width=0.5\textwidth]{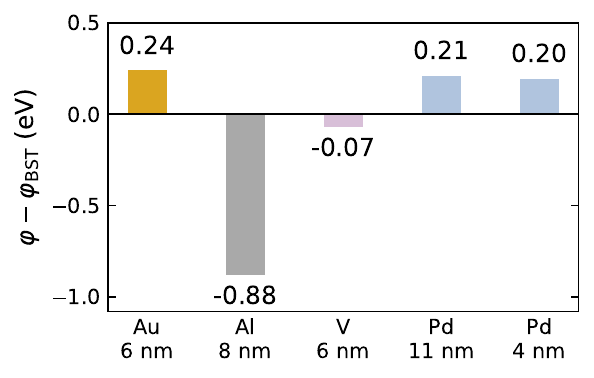}
	\caption{Work function offsets between a BST film and various metals deposited by electron beam evaporation, measured by KPFM. The value plotted is the work function of the metal referenced to the work function of BST. Values for two thicknesses of deposited Pd are shown; when the Pd is thin (4~nm) the KPFM measurement should be sensitive to the Pd-Te alloy; when the Pd is thick (11~nm) the measurement should be sensitive to excess Pd metal sitting atop the film.}
	\label{sfig:kpfm_bar}
\end{figure*}

Kelvin probe force microscopy (KPFM) was used to measure the work function offset between the BST film and deposited metal films. KPFM measures the work function of a surface referenced to a Pt-coated tip, but the absolute value of the measured work function may not be accurate and only the relative work function difference should be considered. A KPFM image of a test junction structure is shown in Fig.~\ref{sfig:kpfm_img}. Work function offsets between BST and a number of deposited metals are shown in Fig.~\ref{sfig:kpfm_bar}.

We find an offset of nearly 900~meV between BST and \ce{Al}. The measured offset between BST and V is much smaller, but this likely is because KPFM is sensitive to vanadium oxide on the surface of the metal (the work function of \ce{VO2} is 5.15~eV~\cite{ko2011VO2}, whereas the work function of \ce{V} is 4.3~eV~\cite{michaelson1977}). The work functions of bulk Au (5.1 to 5.4~eV) and Pd (5.1~eV)~\cite{michaelson1977} are substantially closer to that of \ce{Bi2Se3}-class materials. We found that the work function of Pd-Te (formed by evaporation of 4~nm \ce{Pd}) exceeds the work function of the BST film by roughly 200~meV, a significantly smaller offset than that from elemental superconductors.

All KPFM and accompanying topography data was taken on a Bruker Dimension Icon in PeakForce-KPFM operating mode (frequency-modulated KPFM) with an NSC18 conductive Pt tip. Data was taken at 0.5~Hz scan speed. Potential measurements in Fig.~\ref{sfig:kpfm_img} used 40~nm interleave lift; other measurements used 60~nm interleave lift. Potential was referenced to the tip, not the sample.

\clearpage

\section{Index of devices}

In the main text, we presented data from two Josephson junctions, Devices~1 and 2, formed by the evaporation of 11~nm Pd on BST. In the supplement, we will present data from three additional devices, Devices~3-5. Devices~1-5 were fabricated simultaneously on a single chip.

\begin{figure*}
\centering
	\subfloat{\label{sfig:etched}}
        \subfloat{\label{sfig:unetched}}
        \subfloat{\label{sfig:unetched2}}
        \subfloat{\label{sfig:unetched3}}
	\includegraphics[width=0.95\textwidth]{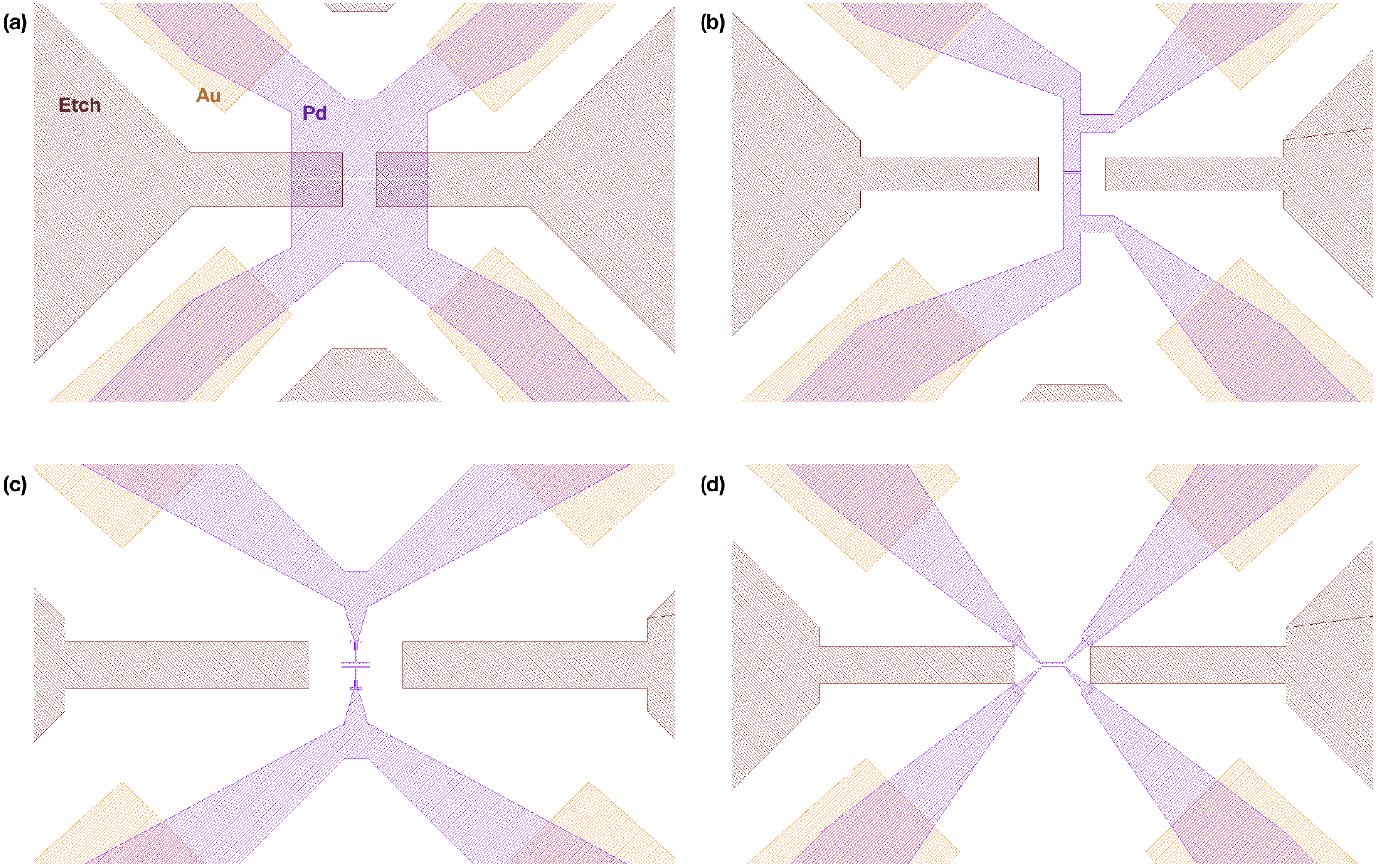}
	\caption{CAD drawings of the device geometries. Initially, the chip is uniformly covered by the BST film. Pd (purple) is then deposited to form the superconducting leads. Unwanted regions of BST are then etched (dark red). Last, Au (orange) is deposited for wirebonding. (a) Etched geometry, used in Devices~1, 3, and 4. (b) Unetched geometry, used in Devices~2 and Al1-4. (c, d) The geometries of Devices~Al5-10 have thin superconducting leads to expel magnetic flux.}
	\label{sfig:CAD}
\end{figure*}

In the supplement, we will also present data from a number of Josephson junctions formed by the evaporation of 5~nm Pd followed by 65~nm Al on BST, which will be labeled Devices~Al1-10. Devices~Al1-4 were fabricated on a second chip, and have similar geometries as Devices~1-5. Devices~Al5-10 were fabricated on a third chip, and have thinner superconducting leads (see Section~\ref{sec:sFraun}). CAD drawings of the different geometries are shown in Fig.~\ref{sfig:CAD}. Note that the length of the junctions presented in the main text ($L=160$~nm) is measured by SEM. This value excludes lateral diffusion of Pd into the weak link (which is not visible in SEM), and is different than the nominal length in the CAD drawings due to electron scattering during electron beam lithography.

\begin{table}[h!]
\centering
\begin{tabular}{ |c|c|c|c|c|c|c| } 
 \hline
 Device & Etched & $I_c$ (nA) & $R_N$ ($\Omega$) & $eI_c R_N/\Delta$ & $eI_c R_N/E_{\rm Th}$ & $eI_e R_N/\Delta$\\ 
 \hline
 1 & Yes & 370 & 146 & 0.30 & 1.2 & 0.11 \\
 2 & No & 640 & 144 & 0.51 & 2.1 & 0.29 \\
 3 & No & 875 & 42 & 0.21 & 0.85 &  \\
 4 & Yes & 580 & 99 & 0.31 & 1.2 &  \\
 5 & No & 375 & 126 & 0.26 & 1.0 &  \\
 6 & Yes & 179 & 301 & 0.30 & 1.2 & 0.20 \\
 \hline
 Al1 & No & 47 & 183 & 0.06 & 0.24 & $-0.03$ \\
 Al2 & Yes & 203 & 79 & 0.10 & 0.40 & $-0.01$ \\
 Al3 & No & 41 & 208 & 0.04 & 0.16 & $-0.03$ \\
 Al4 & No & 95 & 160 & 0.13 & 0.52 & $-0.05$ \\
 \hline
 Al5 & Yes & 263 & 100 & 0.22 & 0.89 & $-0.04$ \\
 Al6 & Yes & 145 & 140 & 0.17 & 0.69 & 0.00 \\
 Al7 & No & 172 & 87 & 0.12 & 0.48 & $-0.01$ \\
 Al8 & No & 120 & 100 & 0.10 & 0.40 & $-0.03$ \\
 Al9 & No & 339 & 72 & 0.20 & 0.81 & $-0.01$ \\
 Al10 & No & 181 & 82 & 0.15 & 0.60 & $-0.03$ \\
 \hline
\end{tabular}
\caption{List of Josephson junction devices measured in this work and properties of their electronic behavior at DC. The second column indicates whether the lateral extent of the junctions was limited by an etched edge. $I_{c,e}R_N$ products are normalized by $\Delta=1.76T_c$ where $T_c$ is the critical current of the leads, not the junction. $I_cR_N$ products are also normalized by the Thouless energy $E_{\rm Th}$ as computed in Section~S2. Blank spaces, insufficient data.}
\label{Tab:Devices}
\end{table}

\clearpage

\section{Theory of Josephson supercurrent}

Real junctions have finite transparency due to backscattering at the SN interface or in the weak link. Let $\tau$ be the overall junction transparency. In a ballistic junction, $\tau=\tau_\text{SN}^2$, where $\tau_\text{SN}$ is the transparency of a single SN interface; in junctions with scattering inside the weak link $\tau$ is smaller. A model for the ABS spectrum of a short junction with finite transparency is~\cite{furusaki1999}:
\begin{equation}\label{eq:AB}
E_B(\phi) = \pm \Delta(T) \sqrt{1-\tau \sin^2\phi}
\end{equation}
where $\Delta(T)$ is the temperature-dependent superconducting gap. For $\tau=1$, the two branches (positive and negative) of the ABS spectrum are degenerate at zero energy when $\phi=\pi$ and $3\pi$. In real junctions, $0<\tau<1$, and the two branches anti-cross at zero energy. The resulting current-phase relationship is $2\pi$-periodic.

The BCS superconducting gap decreases as $T$ approaches $T_c$ and can be modeled by the interpolation:
\begin{equation}\label{eq:deltabcs}
\Delta(T) = \Delta_\text{BCS} \tanh\left(1.74\sqrt{\frac{T_c}{T}-1}\right),
\end{equation}
where $T_c$ is the critical temperature corresponding to the zero-temperature BCS gap $\Delta_\text{BCS}=1.76 k_B T_c$. In SNS-geometry junctions, often there is a region S' of suppressed superconductivity between the S and N regions, with an induced gap $\Delta_\text{ind}<\Delta$ due to the proximity or inverse proximity effect. The temperature dependent induced gap has been expressed implicitly as~\cite{aminov1996, chrestin1997}:
\begin{equation}\label{eq:deltaind}
    \Delta_\text{ind}(T)=\frac{\Delta(T)}{1+\gamma_B\sqrt{\Delta^2(T)-\Delta^2_\text{ind}(T)}/\pi k_B T},
\end{equation}
where $\gamma_B\geq 0$ parameterizes the transparency of the interface between the S and S' regions ($\gamma_B=0$ represents a perfect interface). Equation~(\ref{eq:deltaind}) is most clearly applicable to SNS-geometry junctions with a proximitized region of the normal metal from, for instance, a superconductor deposited on top. In our devices, one might wonder if the region of lateral diffusion of Pd into the weak link serves as an S' region. Lacking clear multiple Andreev reflection (MAR) peaks, however, our data does not provide a feature from which we can fit $\Delta_\text{ind}$ from Eq.~(\ref{eq:deltaind}). Therefore, we will proceed using the simpler BCS temperature dependence of the gap Eq.~(\ref{eq:deltabcs}).

\begin{figure*}
\centering
	\includegraphics[width=0.7\textwidth]{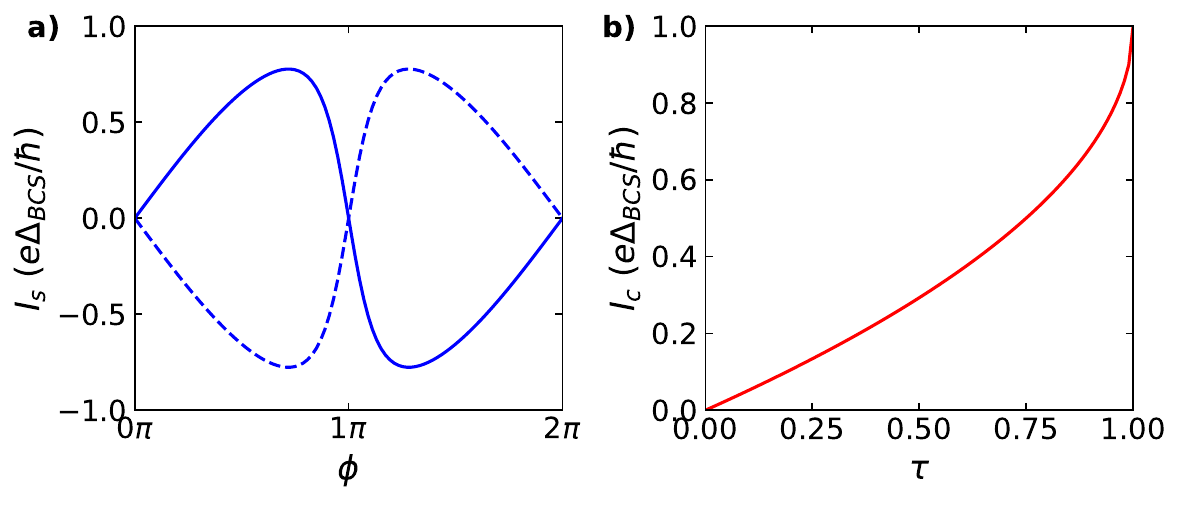}
	\caption{Supercurrent in a model of a short, clean SNS junction at $T=0$. (a) Current-phase relationship due to the ABS spectrum Eq.~(\ref{eq:AB}) at $\tau=0.95$. The solid and dashed lines indicate the two ABS branches. (b) Single-mode critical current Eq.~(\ref{eq:1mode}) as a function of $\tau$.}
	\label{sfig:1mode}
\end{figure*}

In general, the current-phase relationship of SNS junctions non-sinusoidal, so the critical current is not necessarily the current at $\phi=\pi$ and is found by maximizing the supercurrent with respect to $\phi$. In the limit of a short, clean SNS junctions ($L\ll\xi, \ell_e$), the Andreev bound state spectrum is given by Eq.~(\ref{eq:AB}). At zero-temperature, the critical current for a single ABS is~\cite{furusaki1999}:
\begin{equation}\label{eq:1mode}
I_c^1 = \max_\phi \frac{e\Delta}{\hbar}\frac{\tau\sin\phi}{2}\frac{1}{\sqrt{1-\tau\sin^2(\phi/2)}},
\end{equation}
which is $I_c^0\equiv e\Delta/\hbar$ at $\tau=1$ (Fig.~\ref{sfig:1mode}). The critical current of a multi-mode junction is $I_c=N_m I_c^1$, where $N_m$ is the number of transverse modes.

The critical current of Josephson junctions at finite temperature has been calculated in a number of limits. The critical current of SIS-geometry tunnel junctions is given by Ambegaokar-Baratoff theory (AB theory):
\begin{equation}
\frac{I_c(T)R_N}{\Delta_0/e}=\frac{\pi}{2}\frac{\Delta(T)}{\Delta_0}\tanh\frac{\Delta(T)}{2k_BT},
\end{equation}
where the temperature-dependent superconducting gap is given by the BCS expression Eq.~(\ref{eq:deltabcs}).

\begin{figure*}
\centering
	\includegraphics[width=0.75\textwidth]{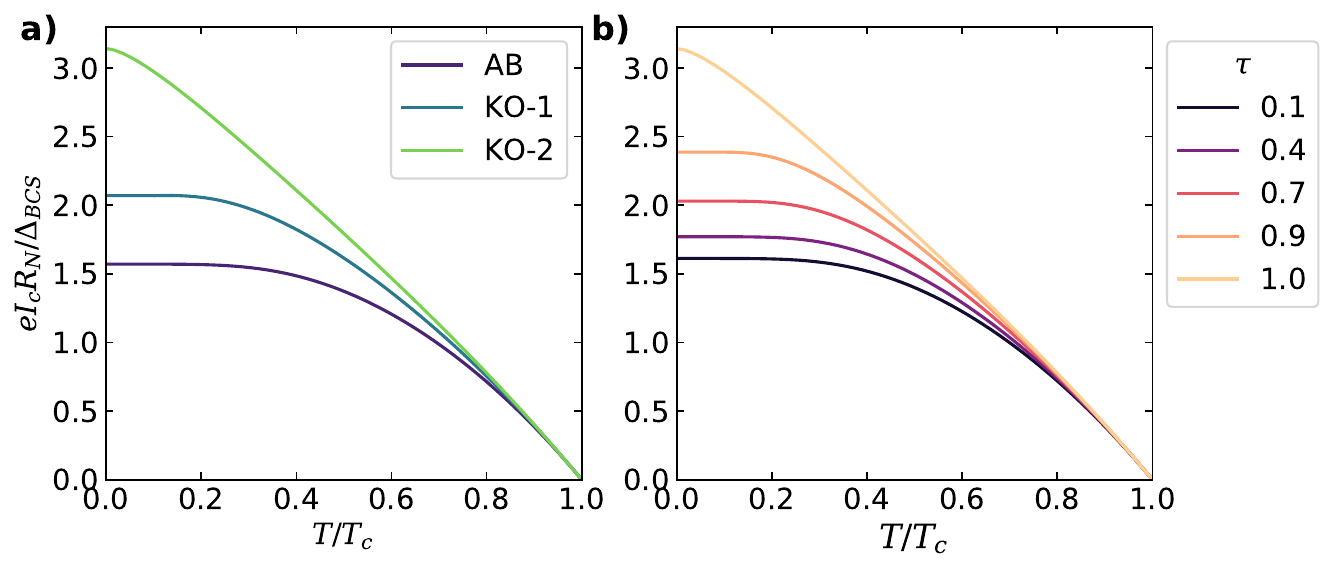}
	\caption{Theoretical value of critical current versus temperature in (a) AB, KO-1, and KO-2 theories, and (b) KO-2 theory with imperfect junction transparencies. The critical current is normalized by the normal-state resistance and the zero-temperature superconducting gap.}
	\label{sfig:KOtheory}
\end{figure*}

The critical current of a short, clean SNS junction is given by Kulik-Omelyanchuk theory of the second type (KO-2 theory) for the case of perfect transmission $\tau=1$~\cite{likharev}:
\begin{equation}
\frac{I_c(T)R_N}{\Delta_0/e} = \max_{\phi} \pi\sin\left(\frac{\phi}{2}\right)\tanh\frac{\Delta(T)\cos(\phi/2)}{2k_BT},
\end{equation}
which has been extended for the case of $\tau\neq1$~\cite{furusaki1999}:
\begin{equation}
\frac{I_c(T)R_N}{\Delta_0/e} = \max_{\phi} \frac{\pi}{2}\frac{\sin{\phi}}{\sqrt{1-\tau\sin^2(\phi/2)}}\tanh\frac{E_B(\phi,T)}{2k_BT}.
\end{equation}
The critical current in the short, dirty junction limit  ($\xi\ll L \ll \ell_e$) is given by Kulik-Omelyanchuk theory of the first type (KO-1 theory)~\cite{likharev}. The temperature-dependent critical current is compared in Fig.~\ref{sfig:KOtheory} between AB theory, KO-1 theory, and KO-2 theory for different values of $\tau$. The value of the normalized critical current $eI_cR_N/\Delta_0$ at zero temperature is $\pi/2$, $\pi$, and 2.07 (at $\tau=1$), respectively. In KO-2 theory, $eI_cR_N/\Delta_0$ drops to $\pi/2$ at low transparency.

\newpage

\section{Junction transparency}

It is important to determine the transparency of the junctions. Beyond serving as a figure of merit describing the quality of the SN interfaces, the transparency determines the amplitude of scattering between forward- and backward-propagating Andreev bound states, and therefore the probability of Landau-Zener tunneling between the bound states. Extracting the transparency of the junctions is not straightforward in our junctions due to the complex structure of their interfaces.

The transparency of short, clean SNS-geometry junctions is modeled by a single effective transparency $\tau\in (0,1)$. The value $\tau$ is approximately the square of the transparency $\tau_\text{SN}$ at each idealized SN interface, less reflections due to backscattering inside the weak link at finite length.

In this section, we present a number of separate tools by which we estimate the transparency. Values are shown in Table~\ref{Tab:Trans} for two junctions with Pd-Te superconducting leads, and one junction with hybrid Pd-Te/Al leads. The most prevalent estimate for junction transparency is the fourth tool we present, the excess current, from which we infer a transparency $\tau$ of roughly 0.3.

\begin{table}[h!]
\centering
\begin{tabular}{ |c|c|c|c| } 
 \hline
 Estimate & Device 1 & Device 2 & Device Al5 \\
 \hline
 1. Normal state $h/e^2N_mR_N$ & $\geq 0.17$ & $\geq 0.16$ & $\geq 0.25$\\
 2. Absolute critical current & 0.06 & 0.09 & $\geq 0.02$ \\
 3. $I_c$ versus $T$ & no data & poor fit & $0.98^*$\\
 4. Absolute excess current & 0.25 & 0.3 & 0.2-0.4\\
 \hline
\end{tabular}
\caption{Summary of estimate of $\tau$ for three devices using various tools. $^*$This value comes from a fit whose quality is likely coincidental.}
\label{Tab:Trans}
\end{table}

\subsection{Transparency from $R_N$}

Finite-width junctions support multiple transverse modes. Using the measured carrier density in the BST, we estimate the number of transverse modes as $N_m=2k_FL=1100$ for $W=2$~\textmu m (Devices~Al5-10 have $W=2.4$~\textmu m, giving $N_m=1310$), where $k_f=\sqrt{\pi n/2}$ is the Fermi wavevector. The factor of two includes modes on both the top and the bottom surfaces.

The resistance of a diffusive weak link in the normal state is given by:
\begin{equation}
R = N_S \rho_{xx} + \frac{h}{\tau N_m e^2},
\end{equation}
where $\rho_{xx}$ is the sheet resistivity of the weak link materials and $N_S = L/W$ is the number of squares. The resistance of a ballistic weak link is the Buttiker contact resistance alone $R = h/\tau N_m e^2$. Our junctions are likely in a quasi-diffusive regime, so the diffusive contribution to $R$ is less than $N_S \rho_{xx}$. Thus the formula for the ballistic case gives a lower bound for $\tau$. For Devices~1 and Al5 we extract lower bounds $\tau=0.17$ and $\tau=0.25$, respectively.

\subsection{Transparency from $I_c(T=0)$}

Here, we present a crude estimate of junction transparency by equating the zero-temperature critical current with the single-mode critical current of a short, clean junction Eq.~(\ref{eq:1mode}) multiplied by the number of transverse modes. We ignore that the devices are closer to the long, dirty junction limit and that there may be an induced gap $\Delta_\text{ind}<\Delta$. For Device~2, using $\Delta_0=1.76k_BT_c$ we have $I_c=0.011N_m I_c^0$, giving $\tau=0.02$. Finite length $L>\xi$, however, suppresses the critical current by the factor $\alpha=1/(1+L/2\xi)$~\cite{bagwell1992}. Taking $L= 160$~nm, we have $L/\xi = 7.1$, giving $\alpha = 0.22$. Accounting for finite length suppression yields $\tau$ = 0.10. For devices with Al, we have no precise estimate of the coherence length, so we present only a lower bound of $\tau$ that does not account for suppression of $I_c$ from finite length.

\subsection{Transparency from $I_c(T)$}

Here, we attempt to use KO theory to determine junction transparency normalized to the normal-state resistance. Whereas KO-1 and KO-2 theory predict $eI_cR_N/\Delta$ in the range $\pi/2$ to $\pi$ at zero temperature, we find substantially lower values $eI_cR_N/\Delta_\text{BCS}=0.30$ (Device~1) and $0.36$ (Device~2).

Mechanisms leading to a reduction in the critical current may include finite length, scattering within the weak link, the Andreev spectrum scaling as the induced gap $\Delta_\text{ind}<\Delta_\text{BCS}$, and the presence of quasiparticles, which cause scattering between ABS branches.

\begin{figure*}
\centering
	\includegraphics[width=.8\textwidth]{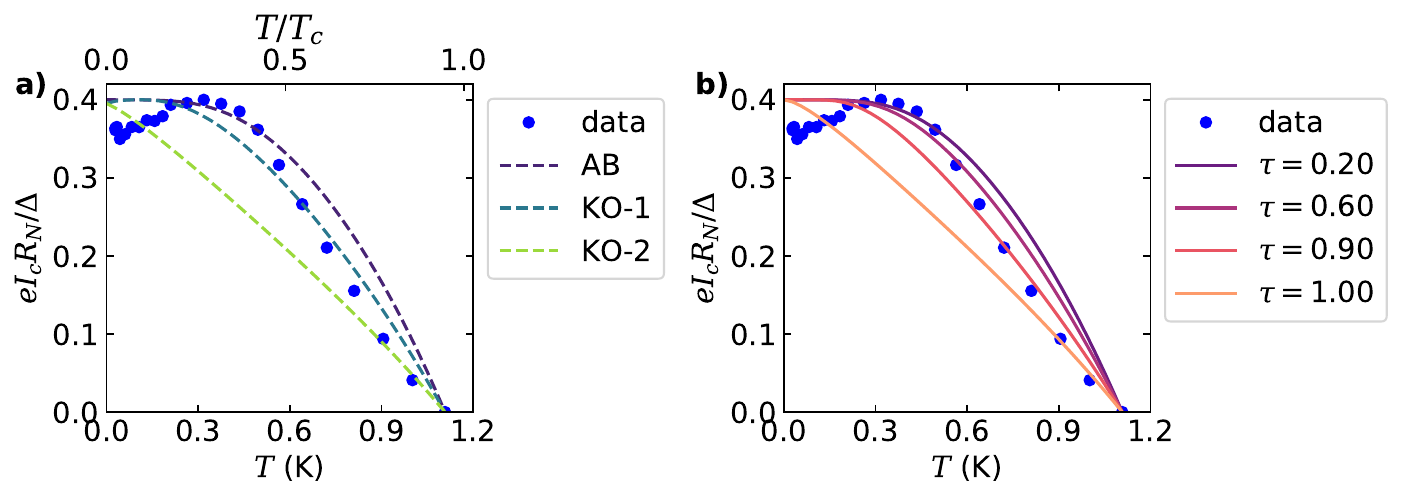}
	\caption{Critical current of Device~2 versus temperature. (a) Dashed lines indicate the expected lineshapes from KO-1, KO-2 ($\tau=1$), and AB theory. (b) Lines indicate the expected KO-2 lineshape at various transparencies.}
	\label{sfig:M13J10_KO}
\end{figure*}

\begin{figure*}
\centering
	\includegraphics[width=.8\textwidth]{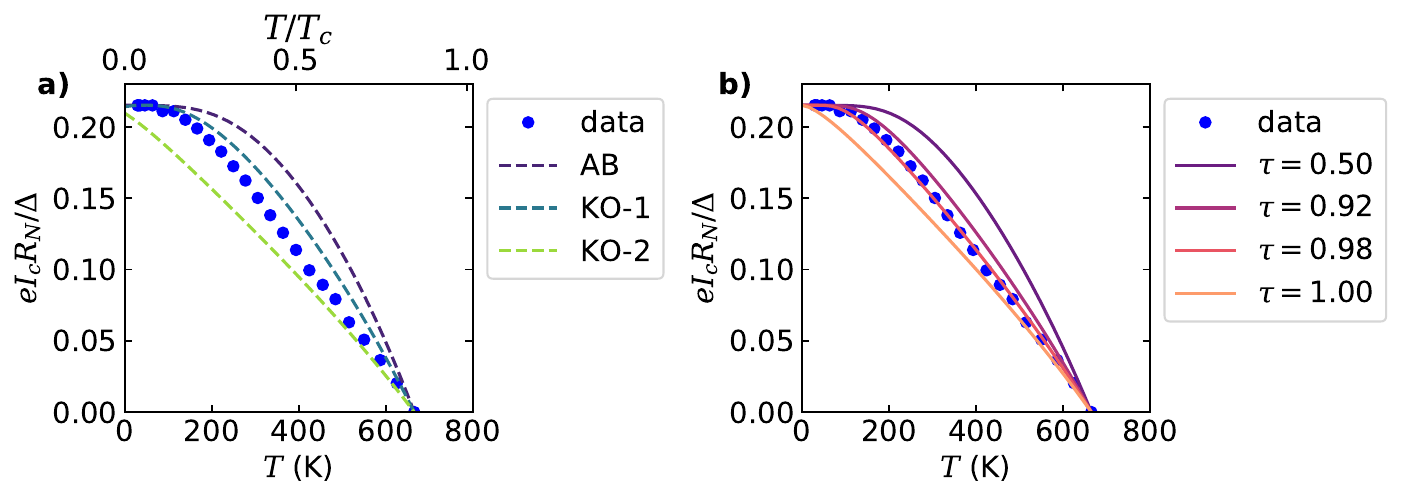}
	\caption{Critical current of Device~Al5 versus temperature (a) Dashed lines indicate the expected lineshapes from KO-1, KO-2 ($\tau=1$), and AB theory. (b) Lines indicate the expected KO-2 lineshape at various transparencies.}
	\label{sfig:V7J1_KO}
\end{figure*}

Ignoring the suppression of the critical current, we can analyze the temperature dependence of the critical current by matching it to KO-1 and KO-2 theories normalized to the measured zero-temperature critical current. The theoretical lineshapes are compared to data from Device~2 in Fig.~\ref{sfig:M13J10_KO}. Note that these are not fits: $T_c$ is taken to be the critical temperature of the junction and critical currents are scaled to the measured $I_c(T=0)$. The theoretical lineshapes do not match the data. As shown in Fig.~\ref{sfig:V7J1_KO}, the temperature dependence of the critical current of Device~Al5 matches the normalized KO-2 lineshape with $\tau \approx 0.98$, however this is likely coincidental. A good fit was not achieved in other devices with a hybrid Pd-Te/Al superconductor.

\subsection{Transparency from $I_e$}

The excess current $I_e$ is the most commonly used measure of transparency of SNS junctions in the experimental literature. The transparency of an SN interface can be extracted from the excess current using BTK theory~\cite{btk}. This model was later extended to include resonances in SNS geometries (OTBK theory)~\cite{OTBK}, however $\tau$ can be extracted in the SNS geometry using BTK theory simply by taking $\tau=\tau_\text{SN}^2$.

In SNS junctions where the weak link N is a conventional metal, the excess current is usually determined by linear extrapolation of the current-voltage relationship in the range $eV=2\Delta_\text{BCS}$ to $3\Delta_\text{BCS}$ to zero voltage. This ensures the exclusion of MAR features. When the weak link is a more complicated system, however, extrapolation of the current-voltage relationship at such high voltages can cause erroneous attribution of non-linearity in the current-voltage relationship of the junction that is unrelated to Andreev reflection. In our system, the resistance of BST has a strong electric field dependence, and extrapolation of the current-voltage relationship at high bias would give an erroneously high excess current and therefore erroneously high transparency.

To provide a dependable estimate for the transparency, we extrapolate the excess current based on the current-voltage relationship slightly beyond the highest resonance feature, typically at around $0.5\Delta_\text{BCS}$. This approach is reasonable as we observe no MAR features. Device~2 has $I_eR_N=0.29\Delta_\text{BCS}/e$, which corresponds to $\tau\approx 0.3$, while Device~Al5 has $I_eR_N=-0.04\Delta_\text{BCS}/e$, corresponding to $\tau\approx 0.2$. However, the excess current of Device~Al5 becomes more negative as $T\rightarrow T_c$ (Fig.~\ref{sfig:exc_tan}), suggesting mechanisms aside from Andreev reflection contribute to the apparent excess current. To isolate the component of excess current arising from Andreev reflection, we suggest considering $\tilde{I}_e=I_e(T=0)-I_e(T_c)=0.06$, which gives $\tau\approx0.4$.

\clearpage

\section{Extended data: current-voltage characteristic and resonances above $I_c$}

\begin{figure*}[h!]
\centering
	\includegraphics[width=.35\textwidth]{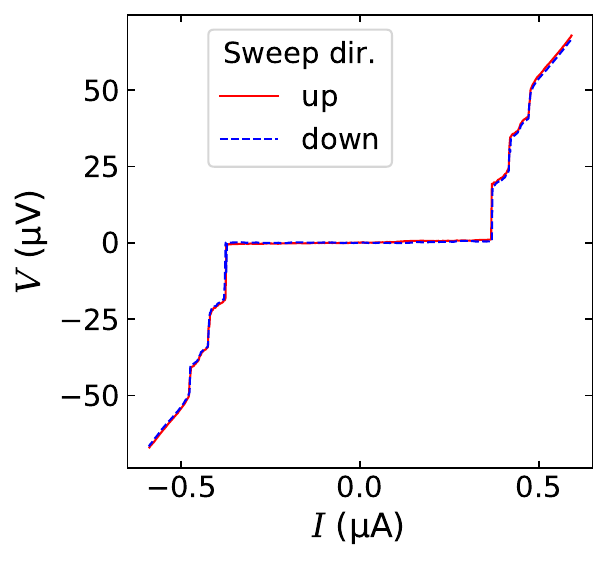}
	\caption{Current-voltage relationship of Device~1 in upwards (red line) and downwards (blue dashed line) sweeps of DC bias. Hysteresis is not observed.}
	\label{sfig:hysteresis}
\end{figure*}

\begin{figure*}
\centering
	\includegraphics[width=.98\textwidth]{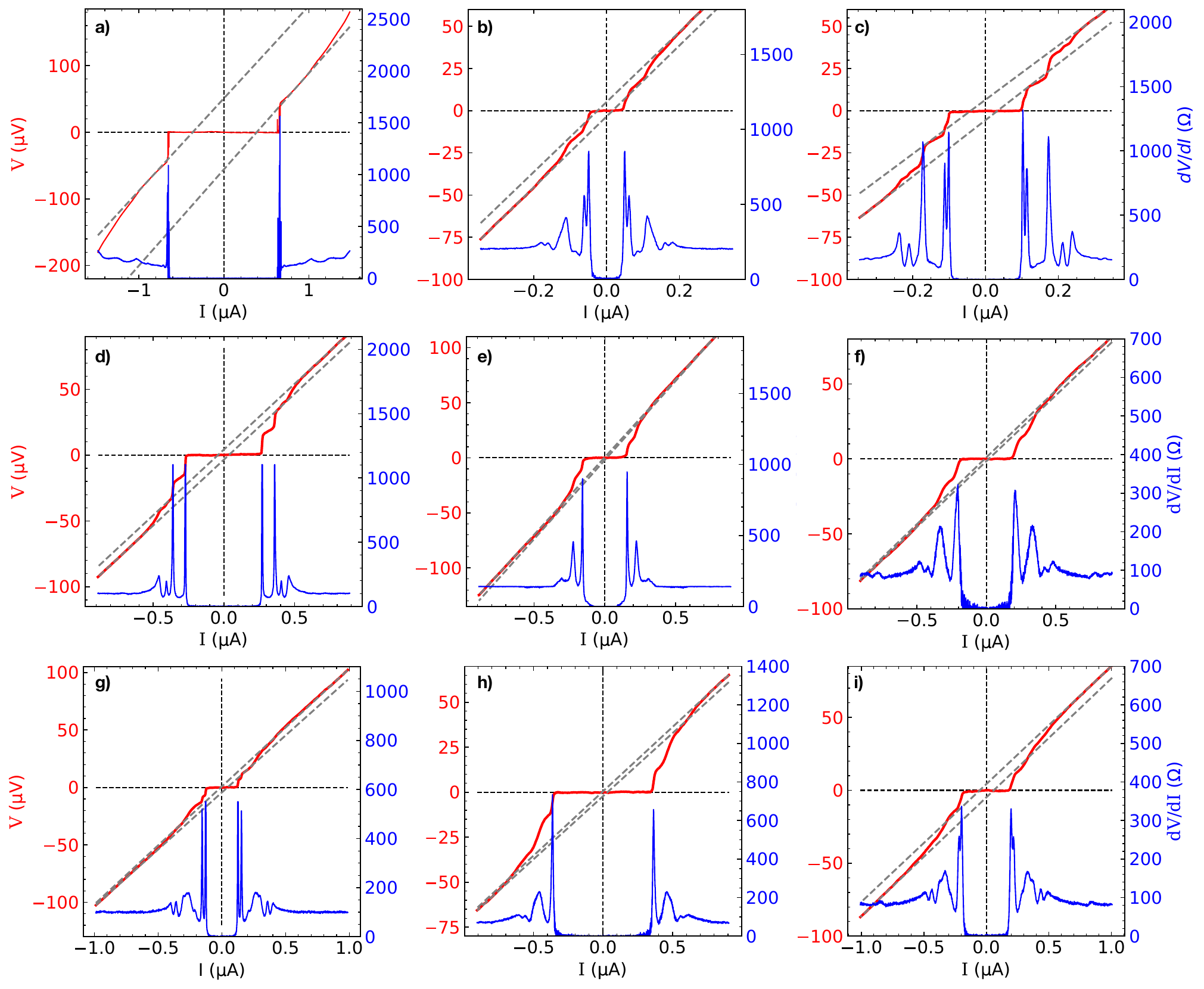}
	\caption{Response of devices under DC current bias. (a) Device~2. (b-i) Devices~Al3-10, respectively. (Red; left axis) DC voltage versus DC current; (blue; right axis) differential resistance. Dashed lines, linear fits used to determine excess current.}
	\label{sfig:bias}
\end{figure*}

The current-voltage relationship of Device~1 under upwards and downwards bias sweeps is shown in Fig.~\ref{sfig:hysteresis}, demonstrating the absence of hysteresis. The current-voltage characteristics of many junctions are shown in Fig.~\ref{sfig:bias}. While a peak in the differential resistance at the critical current $I_c$ is typical, we observe further peaks when $\abs{I}>I_c$.

\begin{figure*}
\centering
	\includegraphics[width=.85\textwidth]{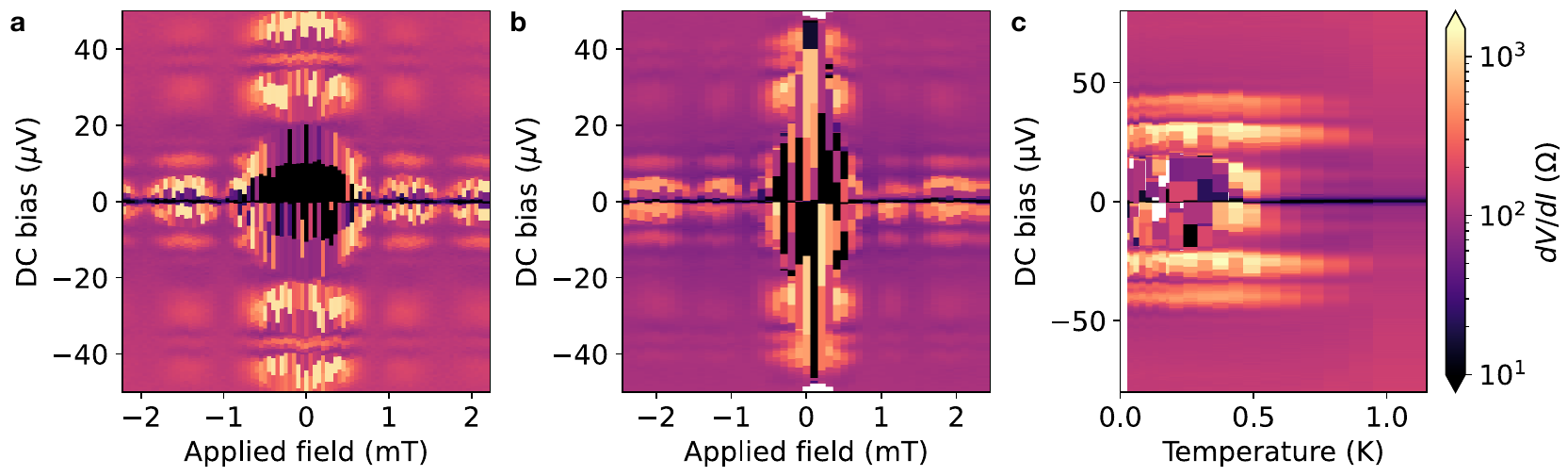}
	\caption{Resonances in the differential resistance, shown as a function of DC bias voltage and applied field or temperature. For clarity, data is shown on a logarithmic scale. (a) Device~1, in magnetic field. (b) Device~2, in magnetic field. (c) Device~2, at elevated temperature. Note that because the impedance of the environment is high compared to that of the devices, data is measured at constant bias current steps. Therefore, when rescaled in voltage units, the data is aliased at the critical temperature and at resonances, where the voltage changes rapidly with small changes in the bias current.}
	\label{sfig:D12_vs_V}
\end{figure*}

\begin{figure*}
\centering
	\includegraphics[width=.7\textwidth]{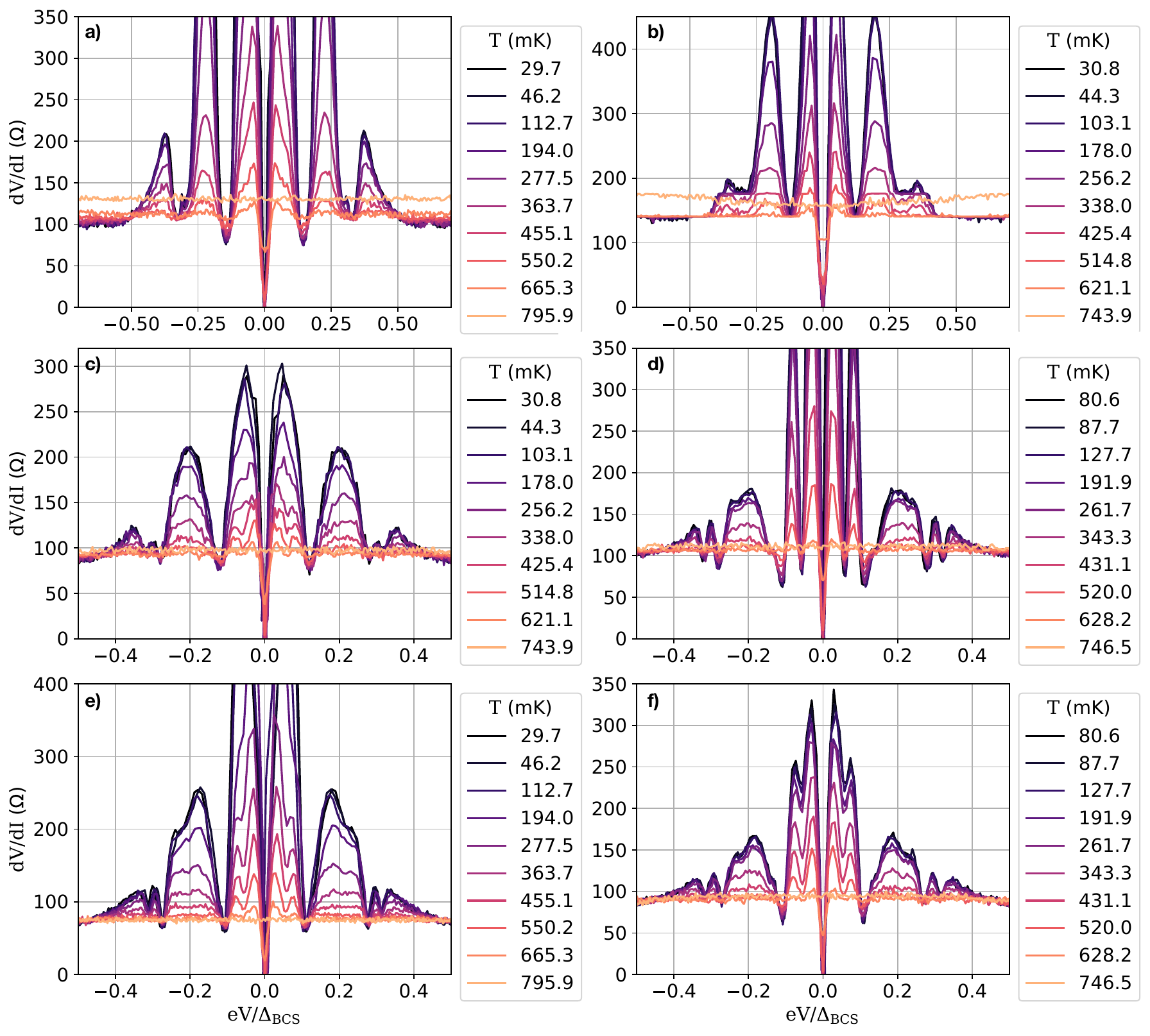}
	\caption{Differential resistance as a function of normalized voltage at different temperatures. a-f) Devices~Al5-10, respectively.}
	\label{sfig:V7heatbias}
\end{figure*}

\begin{figure*}
\centering
	\includegraphics[width=.5\textwidth]{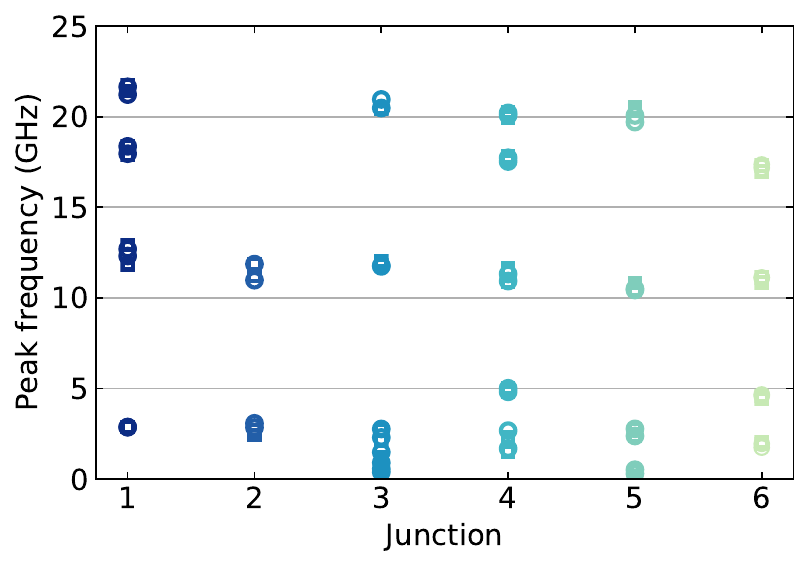}
	\caption{Position of peaks in the differential resistance of Devices~Al5-10, as extracted from current-voltage relationships at base temperature. Peak positions have been converted from voltage to Josephson frequency as $f=2eV/h$. Circles indicate peaks at positive bias ($I>I_c$) and squares indicate peaks at negative bias ($I<-I_c$).}
	\label{sfig:V7_resonances}
\end{figure*}

In Figure~\ref{sfig:D12_vs_V}, resonance features are shown in Devices~1 and 2, with the DC bias rescaled in voltage units, in the presence of applied perpendicular magnetic field and at different temperatures. Figure~\ref{sfig:V7heatbias} shows the response of Devices~Al5-10 to DC bias at different temperatures. While the current bias at which the peaks in differential resistance at $\abs{I}>I_c$ occur shift with applied perpendicular field (e.g. Fig.~ 3(a) of the Main Text), the peaks appear at constant voltage values, as seen in Fig.~\ref{sfig:D12_vs_V}. This is inconsistent with the phenomenology of Fiske steps. Furthermore, the peaks do not shift as $T$ approaches $T_c$, which inconsistent with the phenomenology of coherence peaks from multiple Andreev reflection. The voltages at which peaks appear are compared across Devices~Al5-10 in Fig.~\ref{sfig:V7_resonances}, where the peaks are shown in frequency units. Across devices, peaks appear at a consistent series of frequencies corresponding to centimeter-scale wavelengths, suggesting that the resonances are associated with the cryostat, not the devices. We therefore suspect that the peaks appear due to radiative coupling between device and microwave cavity modes of the cryostat enclosure (the dilution refrigerator's 50~mK plate shield forms a metallic cylindrical cavity around the sample with diameter 5~cm) or to unwanted microwave resonances in the cryostat's low-frequency electrical measurement lines.

\begin{figure*}
\centering
	\includegraphics[width=.75\textwidth]{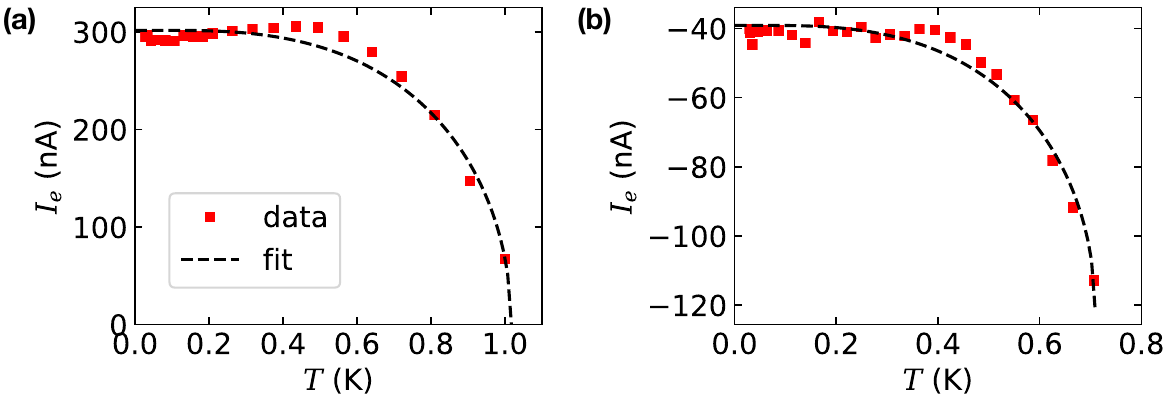}
	\caption{Excess current as a function of temperature in (a) Device~2 and (b) Device~Al5. The data (red dots) are fit by the temperature dependence of the BCS gap Eq.~(\ref{eq:deltabcs}) (dashed lines). The zero-temperature excess current and the size of the gap are used as fitting parameters. In (b), the excess current is increasingly negative with increasing temperature; an offset is added to the fit. No offset is used in (a).} 
	\label{sfig:exc_tan}
\end{figure*}

The excess current is shown as a function of temperature in Fig.~\ref{sfig:exc_tan}. The excess current decreased as $T$ approached $T_c$, and scaled reasonably well with the temperature dependence of the BCS gap Eq.~(\ref{eq:deltabcs}). In Device~2, the excess current scaled with a gap corresponding to a critical current of 1.02~K, whereas $T_c$ of the leads is 1.17~K. Devices with hybrid Pd-Te/Al superconductors often exhibited negative excess current, which became increasingly negative as $T$ approached $T_c$, as if the excess current were offset by an effect unrelated to Andreev reflection. We offer no speculation as to the physical origin of the offset.

\clearpage

\section{Extended data: Fraunhofer patterns and flux jumps}\label{sec:sFraun}

\begin{figure*}[h!]
\centering
	\includegraphics[width=0.4\textwidth]{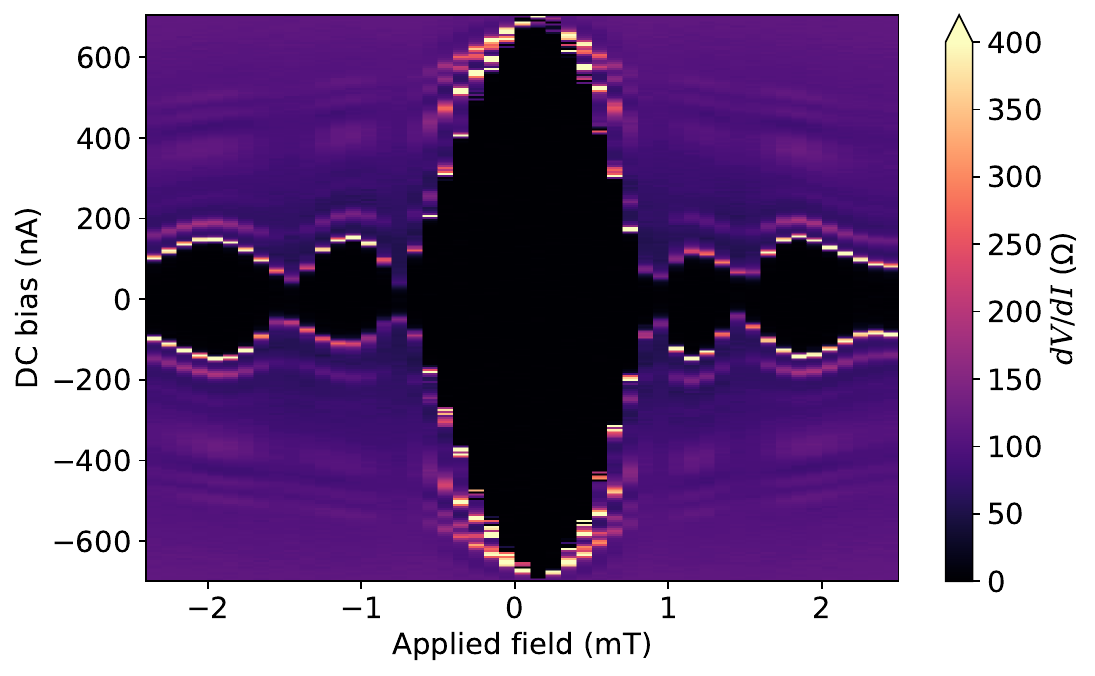}
	\caption{Fraunhofer pattern of Device~2.}
	\label{sfig:Device2Fraun}
\end{figure*}
\begin{figure*}[h]
\centering
	\includegraphics[width=0.8\textwidth]{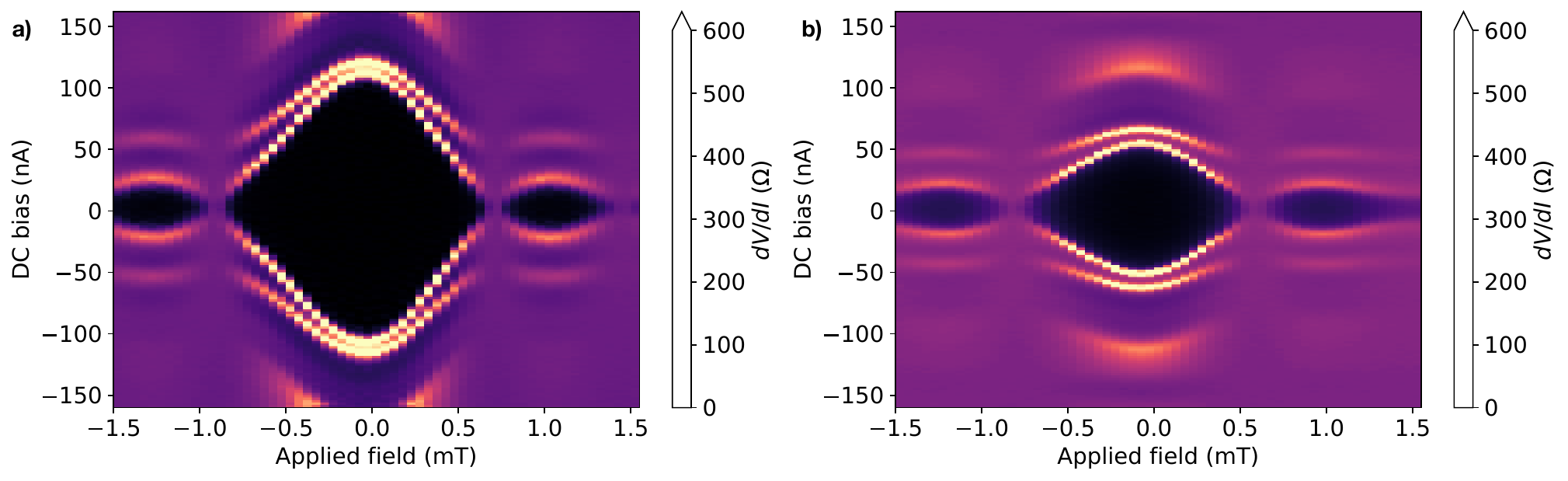}
	\caption{Fraunhofer patterns of Devices (a) Al4 and (b) Al3.}
	\label{sfig:V6fraun}
\end{figure*}
\begin{figure*}
\centering
	\includegraphics[width=0.8\textwidth]{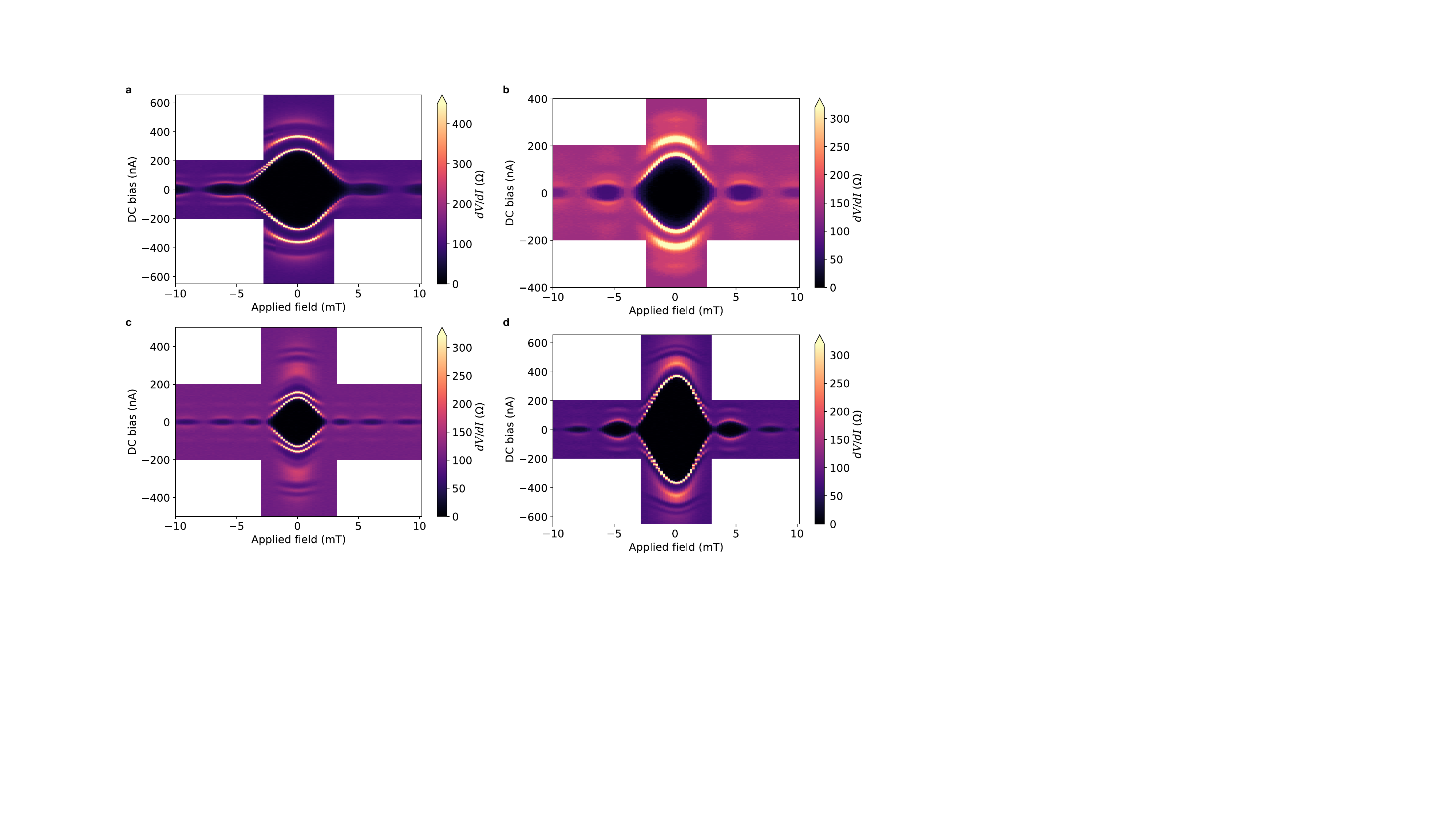}
	\caption{Fraunhofer patterns of Devices (a) Al5, (b) Al6, (c) Al8, and (d) Al9.}
	\label{sfig:V7fraun}
\end{figure*}

The differential resistance as a function of perpendicular field and DC bias is shown for many devices in Figs.~\ref{sfig:Device2Fraun}, \ref{sfig:V6fraun}, and \ref{sfig:V7fraun}. The first nodes in the critical currents of Devices~Al3 and Al4 indicate effective junction areas of roughly 0.7~\textmu m$^2$, while those of Devices~Al5, Al6, Al8, and Al9 indicate areas of roughly 3~\textmu m$^2$.

\begin{figure*}
\centering
	\includegraphics[width=0.8\textwidth]{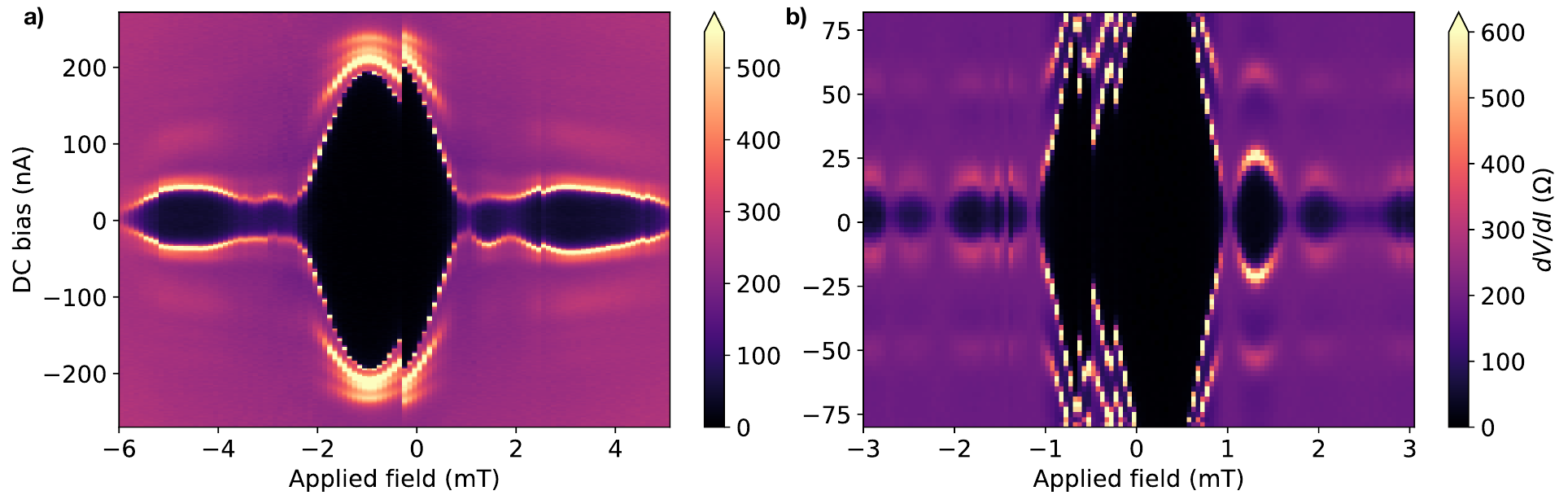}
	\caption{Sudden shifts in the Fraunhofer patterns of (a) Devices~A6 and (b) Al4 are likely flux jumps due to vortices (de)pinning at sites in or near the weak link. Sweep direction is from negative to positive field, and from negative to positive current at each field value.}
	\label{sfig:fluxjumps}
\end{figure*}

Often when sweeping the external field, the critical current suddenly changed. These incidents effectively shifted the Fraunhofer pattern in field. Examples of this behavior in Devices~6 and Al4 are shown in Fig.~\ref{sfig:fluxjumps}. We interpret these shifts as flux jumps, wherein a vortex enters or leaves a pinning site in or near the weak link. Devices~Al5-10 were designed with thin (roughly 150~nm) superconducting leads to make it energetically favorable for flux to be expelled from the leads; no flux jumps were observed in these devices (Fig.~\ref{sfig:V7fraun}). This problem and the same mitigation approach was noted by Ref.~\cite{bai2020}. A problem in Ref.~\cite{bai2020} is that the thin Pd-Te superconductor leads have a critical current near the critical current of the Josephson junction; the addition of \ce{Al} in this work boosts the critical current of the leads well above that of the junction.

\clearpage

\section{Extended data: Shapiro step measurements}

\subsection{Summary}

Shapiro steps were measured in a total of seven devices. Of the four devices with Pd-Te superconducting leads, suppression of the first and third Shapiro steps was observed in one device and suppression of only the first Shapiro step was observed in two devices. Suppressed Shapiro steps were not observed in any of the three devices with hybrid Pd-Te/Al superconducting leads. These results are summarized in Table~\ref{Tab:Shapiro}. Of the devices with $I_C R_N$ products higher than $0.2\Delta/e$, all but one exhibited suppressed odd Shapiro steps, while the devices with lower $I_C R_N$ products expressed all Shapiro steps.

\begin{table}[h]
\centering
\begin{tabular}{ |c|c|c|c|c| } 
 \hline
 Device & Etched & $I_c$ (nA) & $eI_c R_N/\Delta$ & Observation \\ 
 \hline
 2 & No & 640 & 0.51 & Suppressed 1st \& 3rd steps\\ 
 3 & No & 42 & 0.21 & Sup. 1st step; steps at $\sim0.3hf/2e$\\ 
 4 & Yes & 99 & 0.31 & Suppressed 1st step\\ 
 5 & No & 126 & 0.26 & No suppressed steps\\ 
 \hline
 Al1 & No & 47 & 0.06 & No suppressed steps\\ 
 Al2 & Yes & 203 & 0.10 & No suppressed steps\\ 
 Al3 & No & 41 & 0.04 & No suppressed steps\\ 
 \hline
\end{tabular}
\caption{Devices measured under RF irradiation, and a summary of the observed Shapiro step pattern. The second column indicates whether the transverse extent of the junction was defined by etching the TI (Fig.~\ref{sfig:etched}) or whether the TI film was left surrounding the junction (Fig.~\ref{sfig:unetched}).}
\label{Tab:Shapiro}
\end{table}

\clearpage
\subsection{Devices without aluminum}

\begin{figure*}[h!]
\centering
	\includegraphics[width=0.95\textwidth]{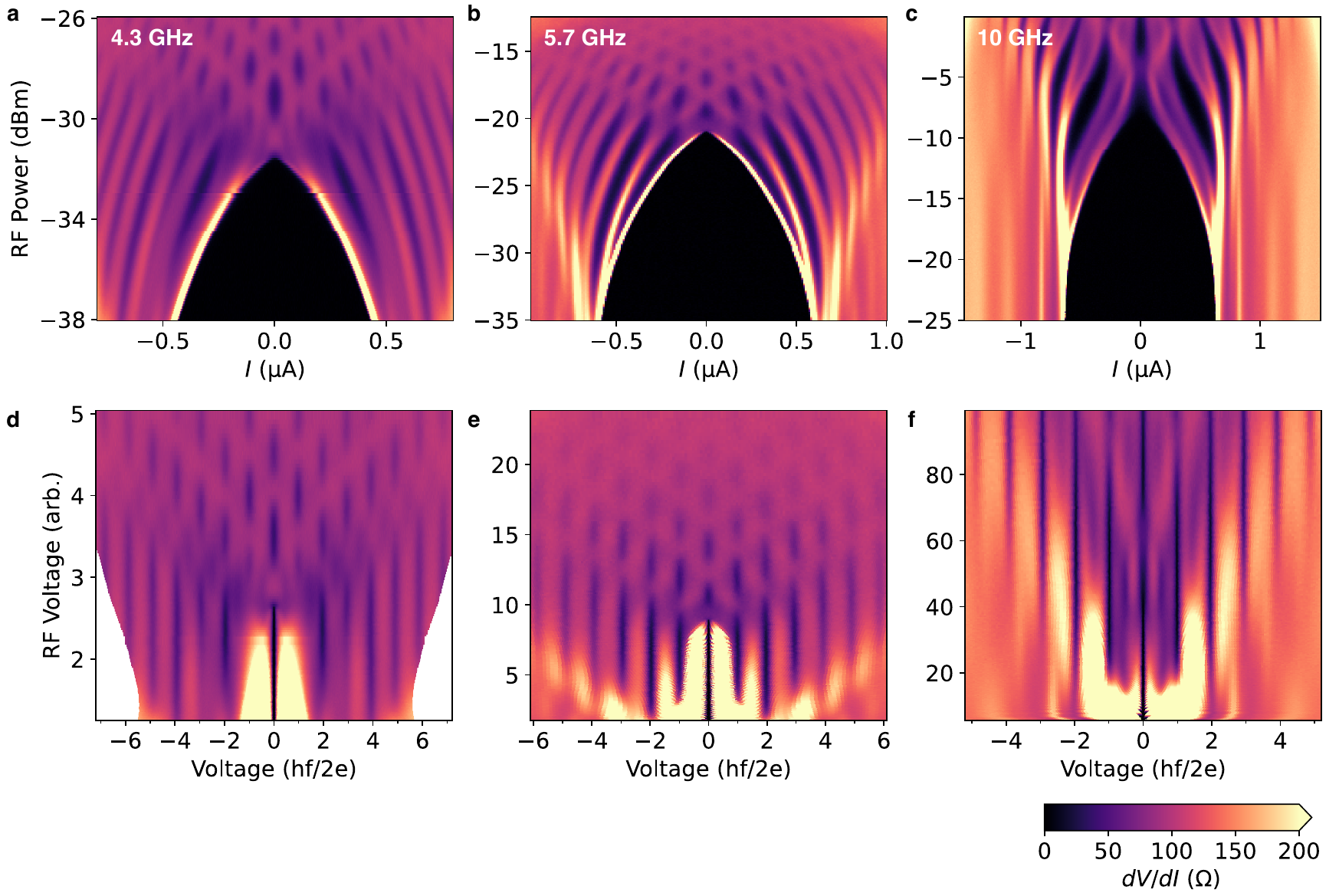}
	\caption{Shapiro steps in Device 2. Data is reproduced from Fig.~5. (a-c) At frequencies 4.3~GHz, 5.7~GHz, and 10~GHz, respectively. (d-f) The same data, shown as a function of DC voltage (x-axis) and relative voltage of the microwave excitation (y-axis). The microwave voltages are shown in arbitrary units because the transmission of the microwave wiring is not known.}
	\label{sfig:M13Shapiro}
\end{figure*}

\begin{figure}
\centering
	\includegraphics[width=0.85\textwidth]{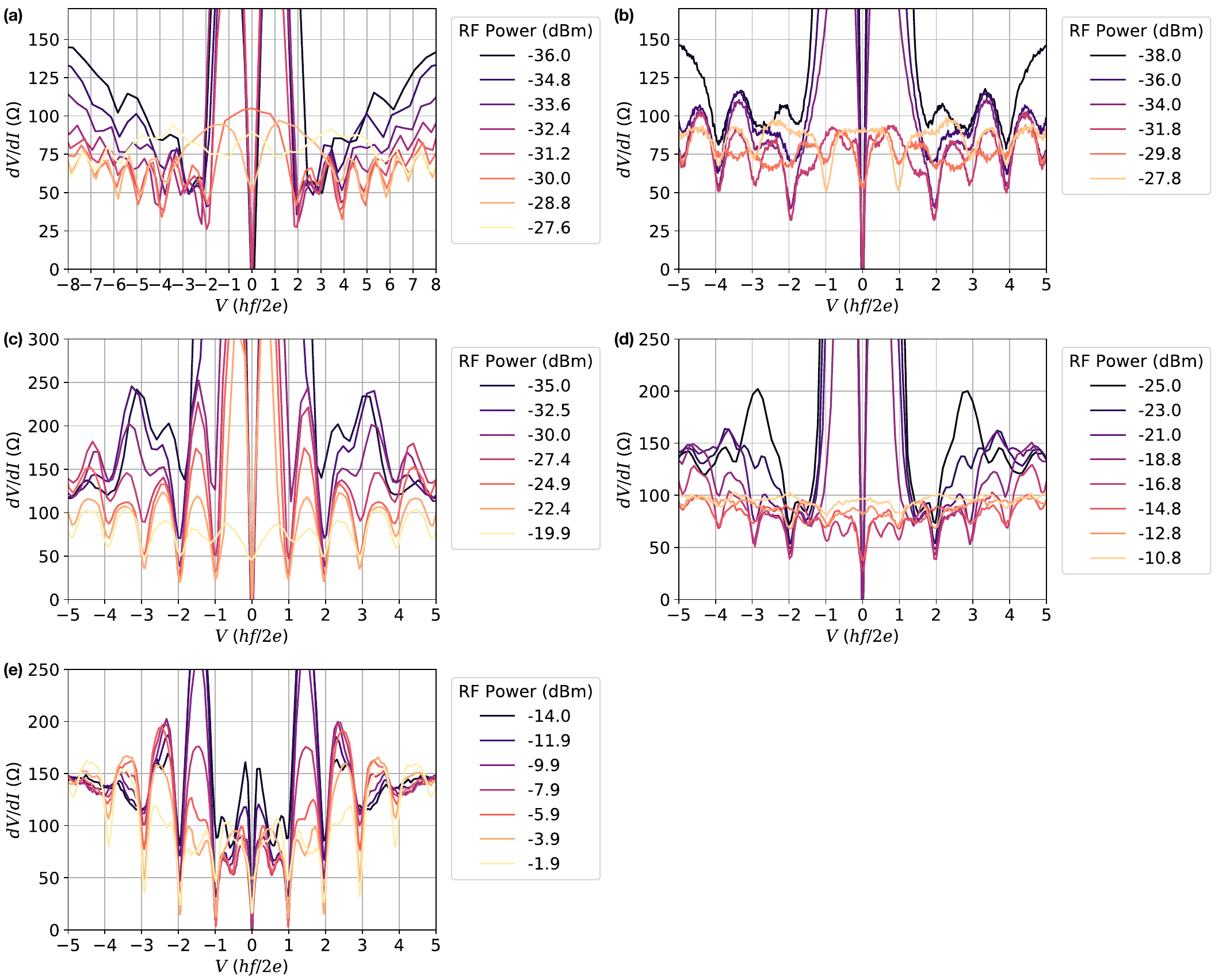}
	\caption{Differential resistance traces versus voltage in Device~2, shown with RF excitation of various power. RF excitation is at frequency (a) 2.5~GHz, (b) 4.3~GHz, (c) 5.7~GHz, (d) 7~GHz, and (e) 10~GHz.}
	\label{sfig:M13Shapiro_cuts}
\end{figure}

The data shown in Fig.~5 of the Main Text (Device~2) are reproduced in Fig.~\ref{sfig:M13Shapiro} with the axes rescaled in units of voltage. Linecuts of these data, along with data at 2.5~GHz and 7~GHz excitation frequencies, are shown in Fig.~\ref{sfig:M13Shapiro_cuts}.

\begin{figure*}[h!]
\centering
	\includegraphics[width=0.95\textwidth]{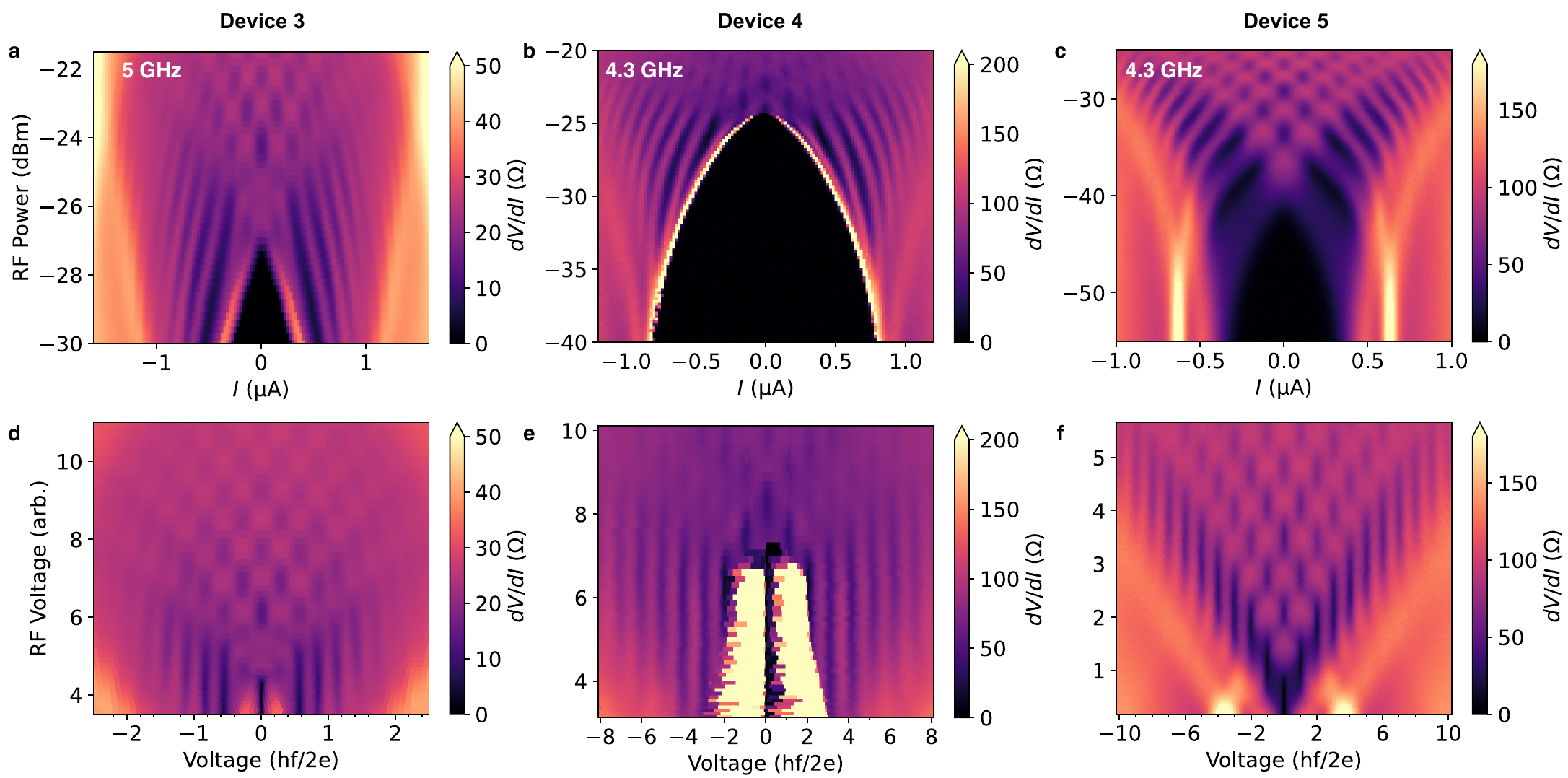}
	\caption{Shapiro steps in (a) Device~3 (at frequency 5~GHz), (b) Device~4 (4.3~GHz), and (c) Device~5 (4.3~GHz). (d-f) The same data, shown as a function of DC voltage (x-axis) and relative voltage of the microwave excitation (y-axis). The microwave voltages are shown in arbitrary units because the transmission of the microwave wiring is not known. Aliasing near the critical voltage, particularly in (e), occurs because the data is sparsely sampled in this region (since data was taken at equally spaced values of current bias, and the DC voltage varies rapidly as a function of current near $I_C$).}
	\label{sfig:Devices345Shapiro}
\end{figure*}

The Shapiro step pattern is shown for devices 3, 4, and 5 in Fig.~\ref{sfig:Devices345Shapiro}. Data is additionally shown with the axes rescaled in units of voltage. Suppression of the first Shapiro step is observed in Devices~3 and 4, but not Device~5. The Shapiro steps in Device~3 occur at integer multiples of approximately $0.3hf/2e$. This observation held at all excitation frequencies tested (2.5, 5, and 7~GHz), and we are unaware of an explanation.

At higher frequencies, half Shapiro steps are expressed, as shown in Fig.~\ref{sfig:M13Shapiro_cuts}(d, e). Note that half Shapiro steps (at multiples of $hf/4e$) are distinct from the fractional AC Josephson effect (where Shapiro steps are expressed at multiples of $hf/e$). Whereas the fractional AC Josephson effect arises from a $4\pi$-periodic current-phase relationship, half Shapiro steps arise from the $\pi$-periodic Fourier component of a $2\pi$-periodic current-phase relationship.

\clearpage
\subsection{Devices with aluminum}

\begin{figure*}[h]
\centering
	\includegraphics[width=0.8\textwidth]{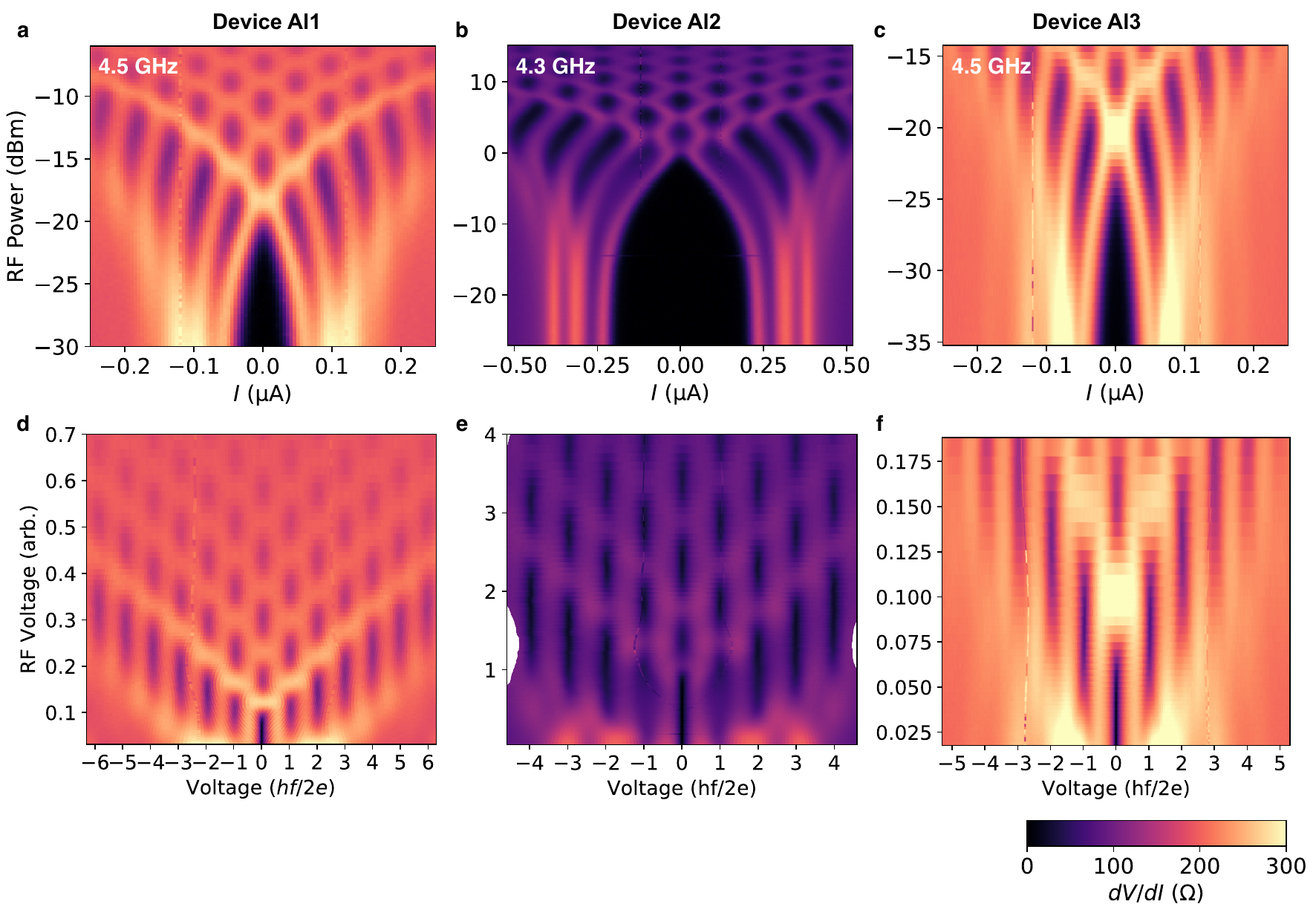}
	\caption{Shapiro steps in devices with hybrid Pd-Te/Al superconducting leads. (a) Device~Al1 (at frequency 4.5~GHz), (b) Device~Al2 (4.3~GHz), and (c) Device~Al3 (4.5~GHz). (d-f) The same data, shown as a function of DC voltage (x-axis) and relative voltage of the microwave excitation (y-axis).}
	\label{sfig:V6Shapiro}
\end{figure*}

\begin{figure*}
\centering
	\includegraphics[width=0.85\textwidth]{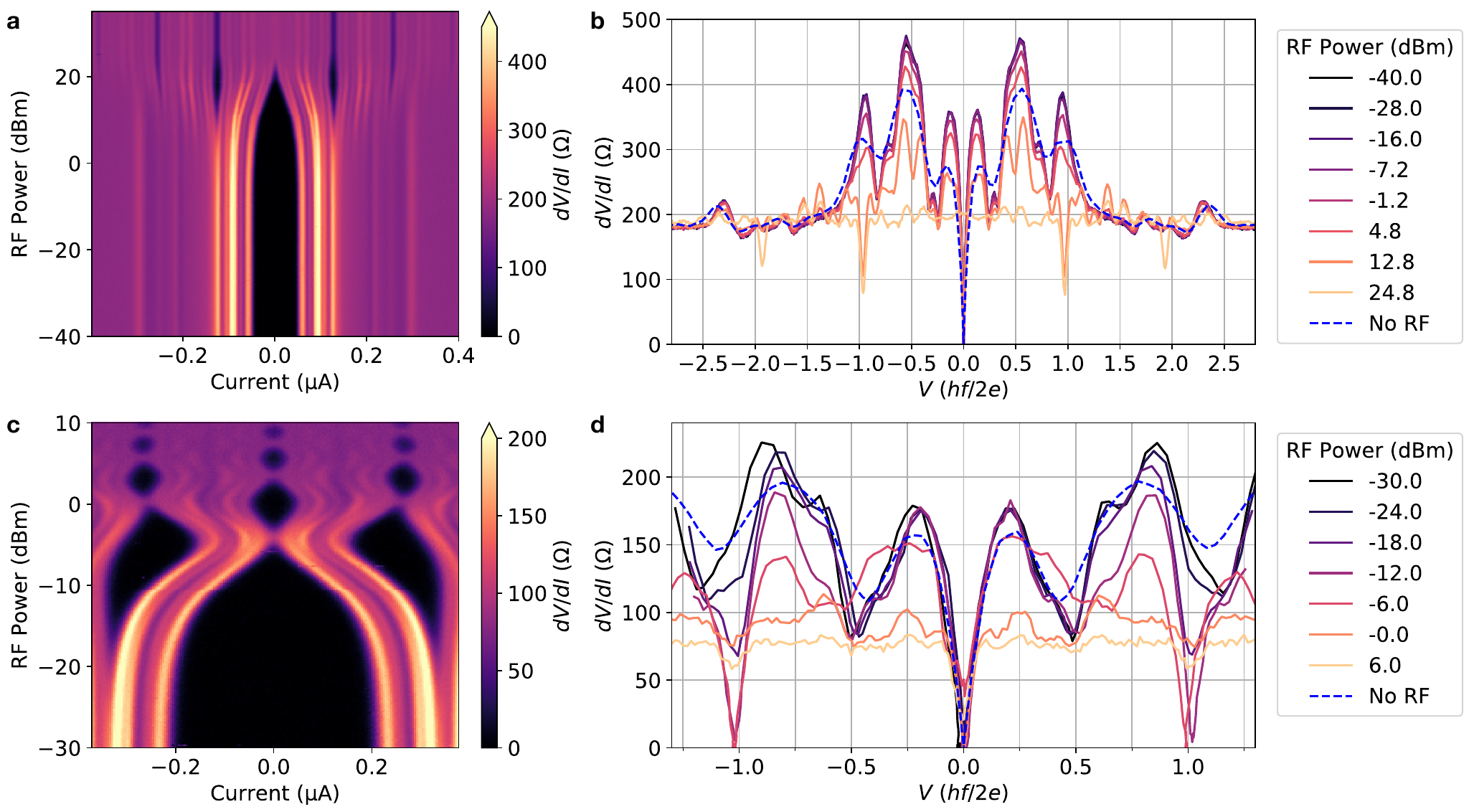}
	\caption{Fractional Shapiro steps under high frequency irradiation. (a) Differential resistance of Device~Al1 at frequency 12~GHz, shown versus current bias and RF excitation power. (b) Linecuts of the data, expressed as a function of bias voltage. The differential voltage without RF excitation (but measured in the same cryostat) is shown by the blue dashed line. (c) Differential resistance of Device~Al2 at frequency 9.7~GHz. (d) Corresponding linecuts.}
	\label{sfig:high_frequency}
\end{figure*}

The Shapiro step pattern is shown for devices Al1, Al2, and Al3, which feature hybrid Pd-Te/Al superconducting leads, in Fig.~\ref{sfig:V6Shapiro}. All Shapiro steps are expressed. At higher frequencies, fractional Shapiro steps emerge, as shown in Fig.~\ref{sfig:high_frequency}. Half Shapiro steps (at odd integer multiples of $hf/4e$) are most prominent, and were also seen in Device~2 (main text Fig.~5(c)). Features at other fractions of $hf/2e$ are present but more difficult to distinguish. It is possible that these features are related to the resonances observed in the devices in the absence of RF excitation; future work on this point is needed.

\clearpage

\section{RSJ simulations of Shapiro steps}

\subsection{Summary}

In this section, we present numerical simulations of the Devices under microwave irradiation. As one would expect, we find that as the $4\pi$-periodic contribution to the total critical current increases, the suppression of odd Shapiro steps increases. The suppression changes the order that Shapiro steps emerge as the microwave power increases. For a $2\pi$-periodic supercurrent, Shapiro steps emerge at sequentially increasing microwave powers. As the $4\pi$-periodic contribution increases, the first signature is that the second step emerges at a lower power than the first step. As the $4\pi$-periodic contribution continues to increase, the fourth step emerges before the third (and eventually before the first), and so on. In addition, as the $4\pi$-periodic contribution grows, the widths of odd Shapiro steps decrease and the widths of even Shapiro steps grow.

\begin{figure*}[h]
\centering
	\includegraphics[width=0.98\textwidth]{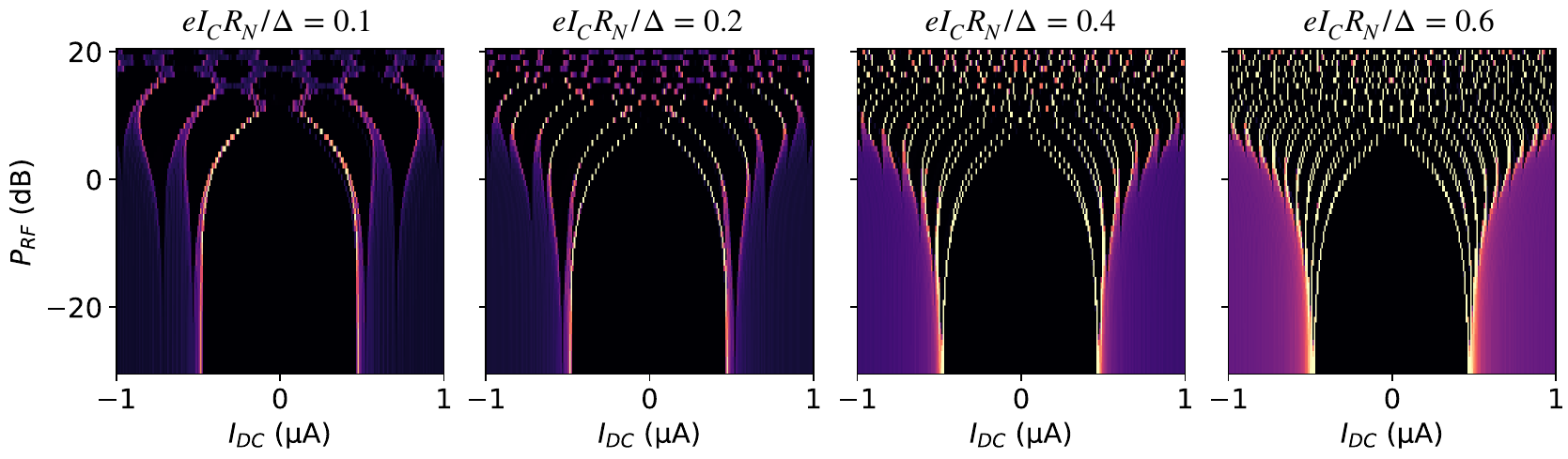}
	\caption{Simulated Shapiro steps in a hypothetical device as $I_C R_N$ varies. We fix $I_C = 500$~nA, $I_{4\pi}=0.1I_C$, and $f=5$~GHz, and we vary $R_N$. As $I_C R_N$ increases, the suppression of odd Shapiro steps becomes increasingly visible.}
	\label{sfig:sim_summary}
\end{figure*}

Less intuitively, we also find that the suppression of odd Shapiro steps increases as the $I_C R_N$ product of the junction increases. To demonstrate this property, in Fig.~\ref{sfig:sim_summary} we present simulations of a hypothetical device with fixed critical current comprising a fixed $4\pi$-periodic contribution, but with variable $R_N$. The suppression of odd Shapiro steps is not visible when $R_N$ is low, and becomes increasingly visible as $R_N$ (and thereby $I_C R_N$) increases.

In the remainder of this section, we present simulation of all devices that we measured under microwave irradiation, using the measured $I_C$ and $R_N$ values of each device at various values of the $4\pi$-periodic component. Among the devices, Device~2 has the highest $I_C R_N$ product ($0.51 \Delta/e$), and we observed suppression of the first and third Shapiro steps. Three more devices have intermediate $I_C R_N$ products around a quarter of $\Delta/e$, and in two of them (Devices~3 and 4, but not 5) we observed suppression of the first Shapiro steps. For none of the devices with lower $I_C R_N$ was suppression of odd Shapiro steps visible. For all but Device~5, the suppression of odd Shapiro steps qualitatively matches simulations when the $4\pi$-periodic component is a few percent of the total supercurrent.

\clearpage
\subsection{Description of simulations}

We numerically simulate time dynamics of the junction according to
\begin{equation}
\frac{\hbar \dot \phi}{2e R_N} + I_{2\pi}\sin(\phi) + I_{4\pi}\sin(\frac{\phi}{2}) = I_\mathrm{DC} + I_\mathrm{RF}\sin(ft),
\end{equation}
using the Runge-Kutta method, and evaluate the instantaneous voltage at each time step as $V=\hbar \dot \phi/2e$. Simulations use roughly 270 time steps per period $1/f$. The DC voltage is determined as the average of the instantaneous voltage throughout 30~periods, but ignoring the first 5~simulated periods. In the simulations, the DC bias applied across the device is presented in power units. Because the effective impedance of the device at microwave frequencies is not known, the power cannot be converted directly to the current bias across the device. Therefore, in simulations, the power of microwave irradiation $P_\text{RF}$ is presented in decibel units with arbitrary reference, and the range of powers displayed is chosen qualitatively in order to reflect the development of the lowest few Shapiro steps. The conversion used in figures is
\begin{equation}
I_\text{DC}=\left( 50\, \Omega \times\, 1 \text{ fW}\, \times\, 10^{\frac{P_\text{RF}}{10}} \right)^{1/2} ,
\end{equation} 
i.e., $P_\text{RF}$ is the power in decibels referenced to 1~fW given 50~$\Omega$ impedance.

\clearpage
\subsection{Simulations of devices without Al}

\begin{figure*}[h]
\centering
	\includegraphics[width=0.9\textwidth]{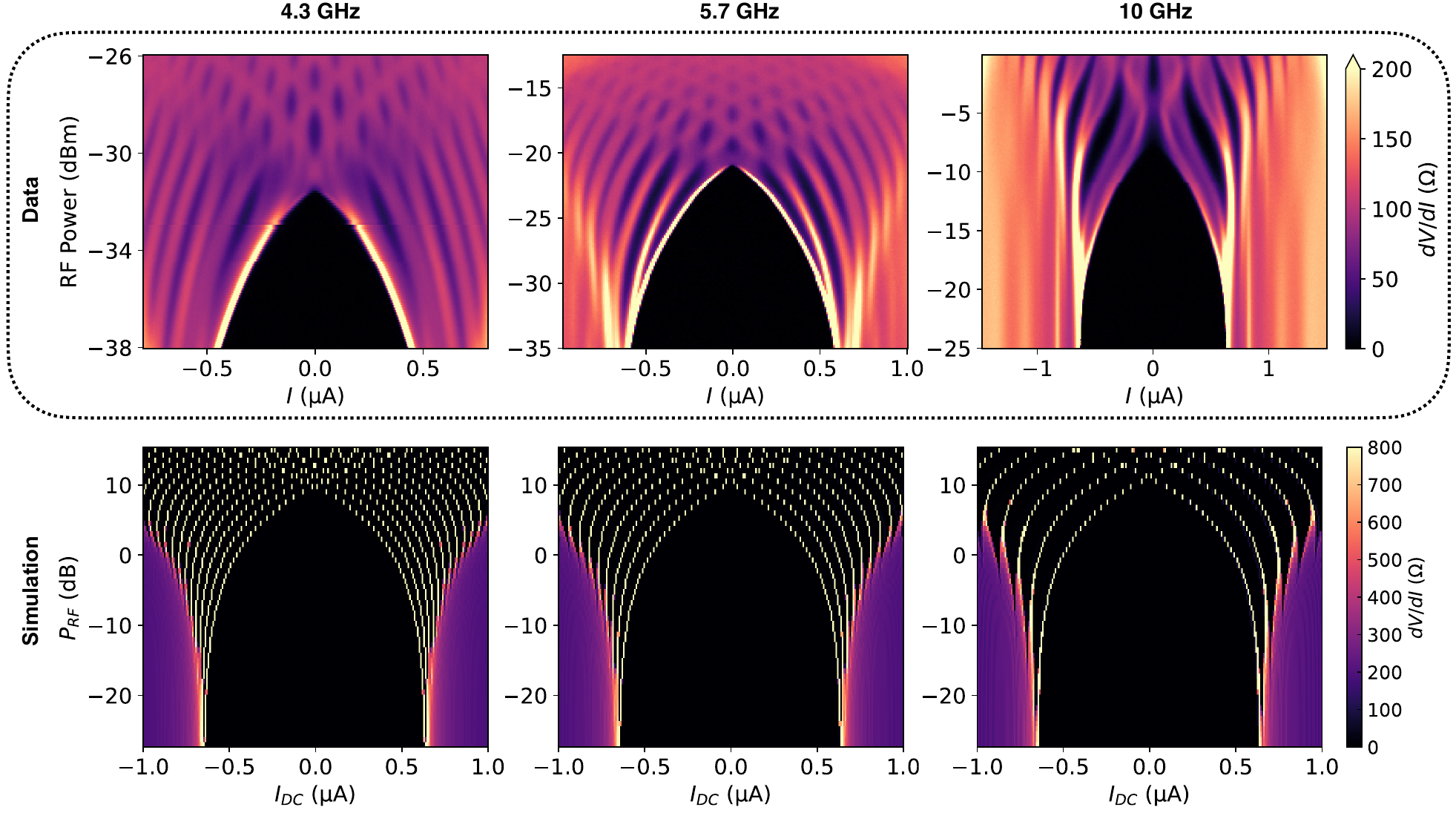}
	\caption{Simulated Shapiro steps in Device~2 at different frequencies, with no $4\pi$-periodic component to the critical current. Experimental data is shown at top for reference. Note the experimental data and simulated data use separate color scales.}
	\label{sfig:device2_sim_vs_freq}
\end{figure*}

\begin{figure*}
\centering
	\includegraphics[width=0.9\textwidth]{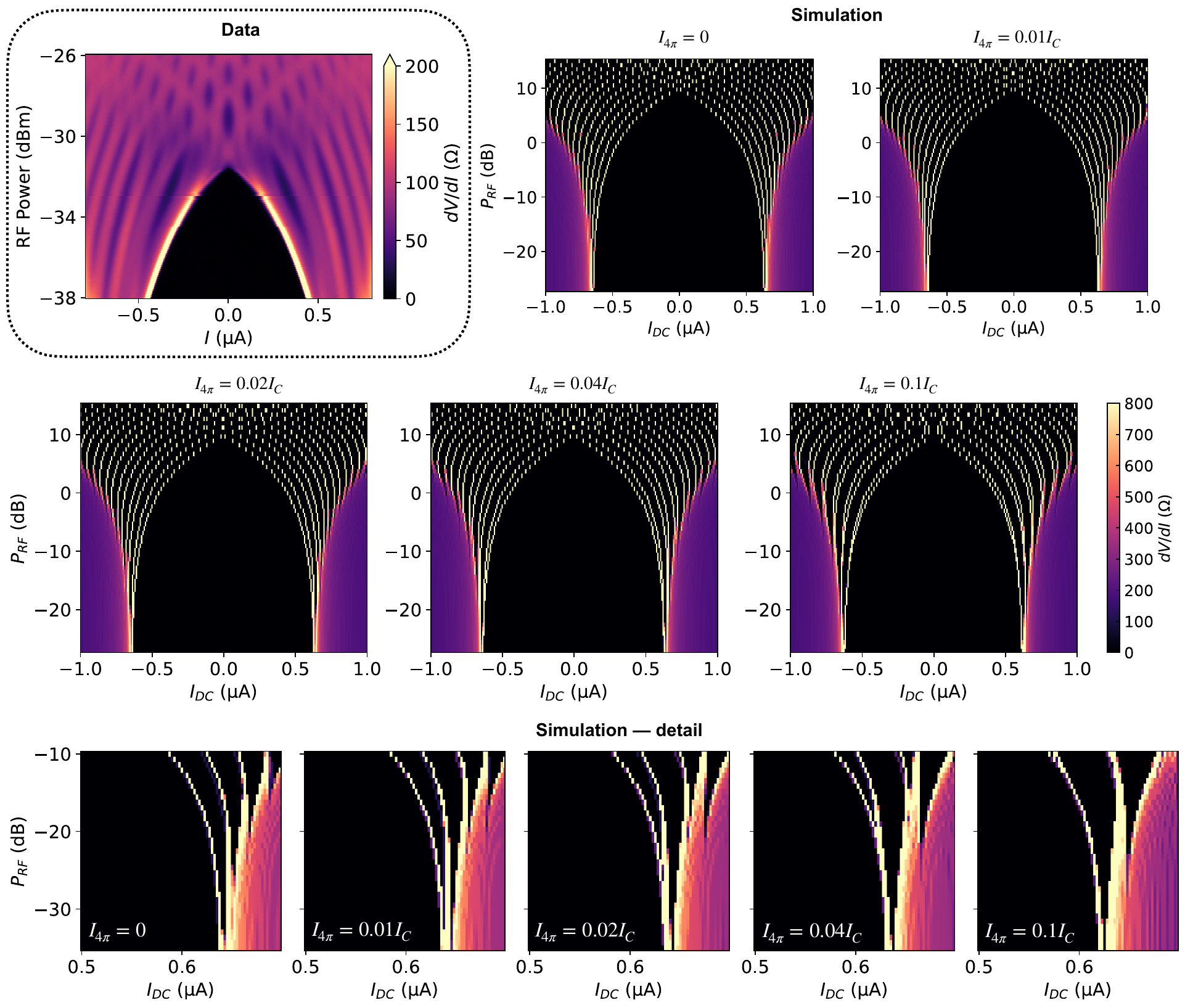}
	\caption{Simulated Shapiro steps in Device~2 at $f=4.3$~GHz, at different ratios of the $4\pi$-periodic critical current. Experimental data is shown at top left for reference, and the onset of the first few Shapiro steps is shown in detailed at bottom. Note the experimental data and simulated data use separate color scales.}
	\label{sfig:device2_sim}
\end{figure*}

Simulations of Device~2 using a $2\pi$-periodic current-phase relationship are presented in Fig.~\ref{sfig:device2_sim_vs_freq}. Because the reduced frequency $\Omega=hf/2eI_C R_N\ll 1$, the simulations feature sharp transitions between Shapiro steps, unlike in the data, where the transitions are gradual. Simulations at $4.3$~GHz with finite $4\pi$-periodicity, shown in Fig.~\ref{sfig:device2_sim}, feature significant suppression of odd Shapiro steps even when the $4\pi$-periodic component is only a few percent of the total critical current. At $I_{4\pi}=0.02I_C$, the second and fourth Shapiro steps develop at lower microwave power than the first and third steps, respectively, and at $I_{4\pi}=0.04I_C$, odd Shapiro steps persist through narrower ranges of DC current as compared to adjacent even steps.

Simulations deviate from measurements, however, in that simulations feature sudden transitions between Shapiro steps as $I_\mathrm{DC}$ varies---a characteristic of Devices irradiated at reduced frequency $\Omega=hf/2eI_C R_N\ll 1$~\cite{park2021}---whereas the data feature gradual transitions ($\Omega=0.097$ for Device~2 at $f=4.3$~GHz). 

\begin{figure*}
\centering
	\includegraphics[width=0.85\textwidth]{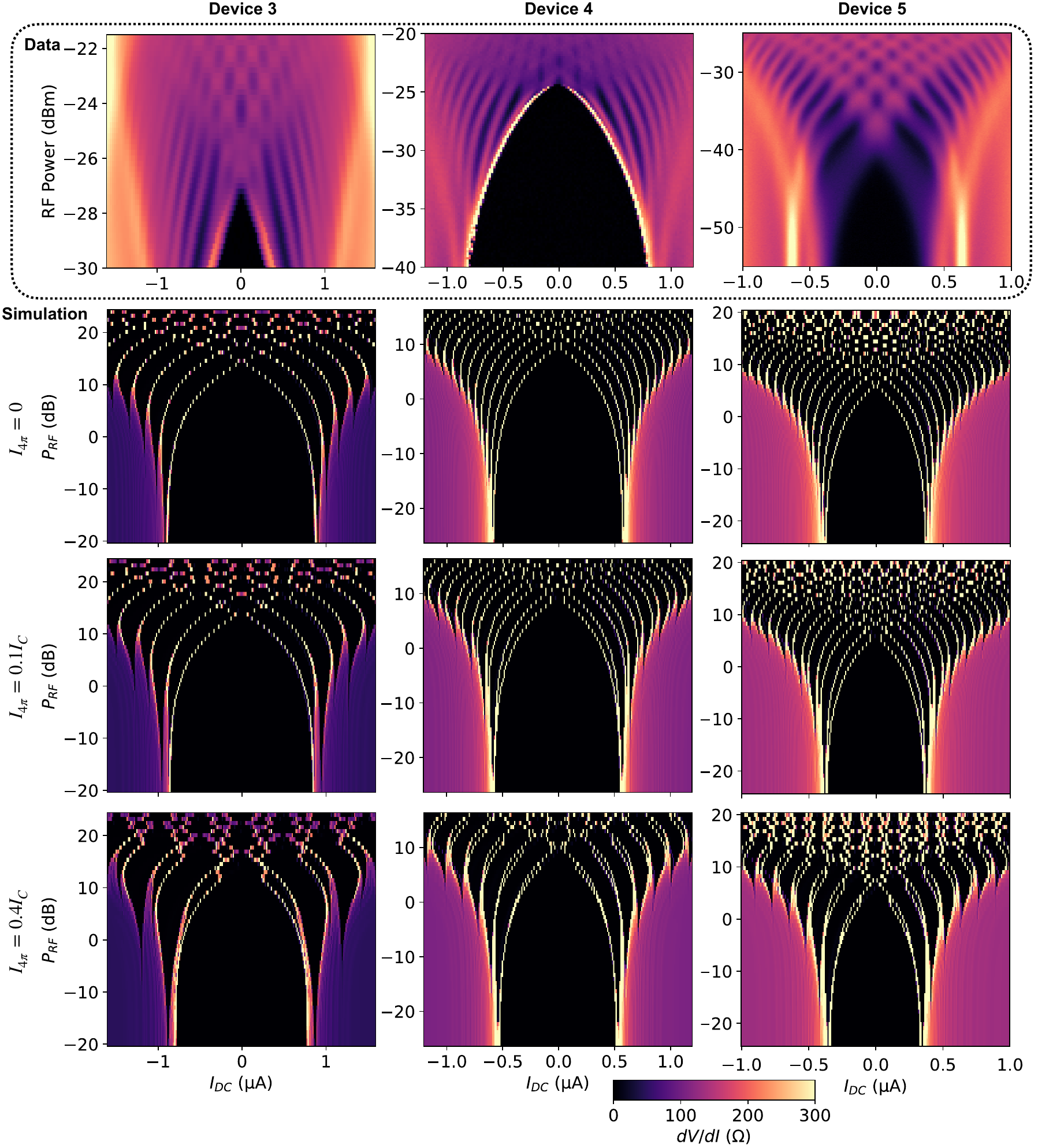}
	\caption{Simulated Shapiro steps in Devices~3, 4, and 5 at different ratios of the $4\pi$-periodic critical current. Experimental data is shown at top for reference.}
	\label{sfig:device345_sim}
\end{figure*}

Simulations of Devices~3, 4, and 5, presented in Fig.~\ref{sfig:device345_sim}, exhibit similar features as simulations of Device~2. Because Devices~3, 4, and 5 have lower $I_C R_N$ products (see Table~\ref{Tab:Devices}), higher $4\pi$-periodic components closer to $10\%$ of the total critical current are required for noticeable suppression of odd Shapiro steps.

\begin{figure*}
\centering
	\includegraphics[width=0.9\textwidth]{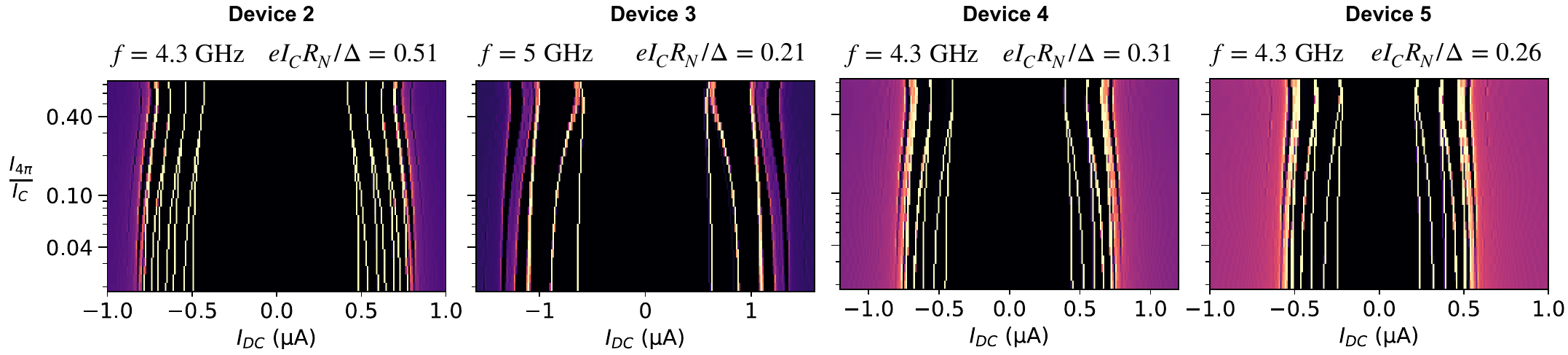}
	\caption{Simulated Shapiro steps in Devices~2-5 as the $4\pi$-periodic critical current is varied. Simulations use constant power: $-2$~dB for Devices~2, 4, and 5, and $5$~dB for Device~3.}
	\label{sfig:sim_vs_4pi}
\end{figure*}

To summarize our findings, in Fig.~\ref{sfig:sim_vs_4pi} we present simulations of the Shapiro steps as a function of the $I_{4\pi}$ components (rather than as a function of microwave power). These results clarify that for a given value of $I_{4\pi}$, the suppression of odd Shapiro steps is most readily visible in Device~2 due to its larger value of $I_C R_N$.

\clearpage
\subsection{Simulations of devices with Al}

\begin{figure*}[h]
\centering
	\includegraphics[width=0.85\textwidth]{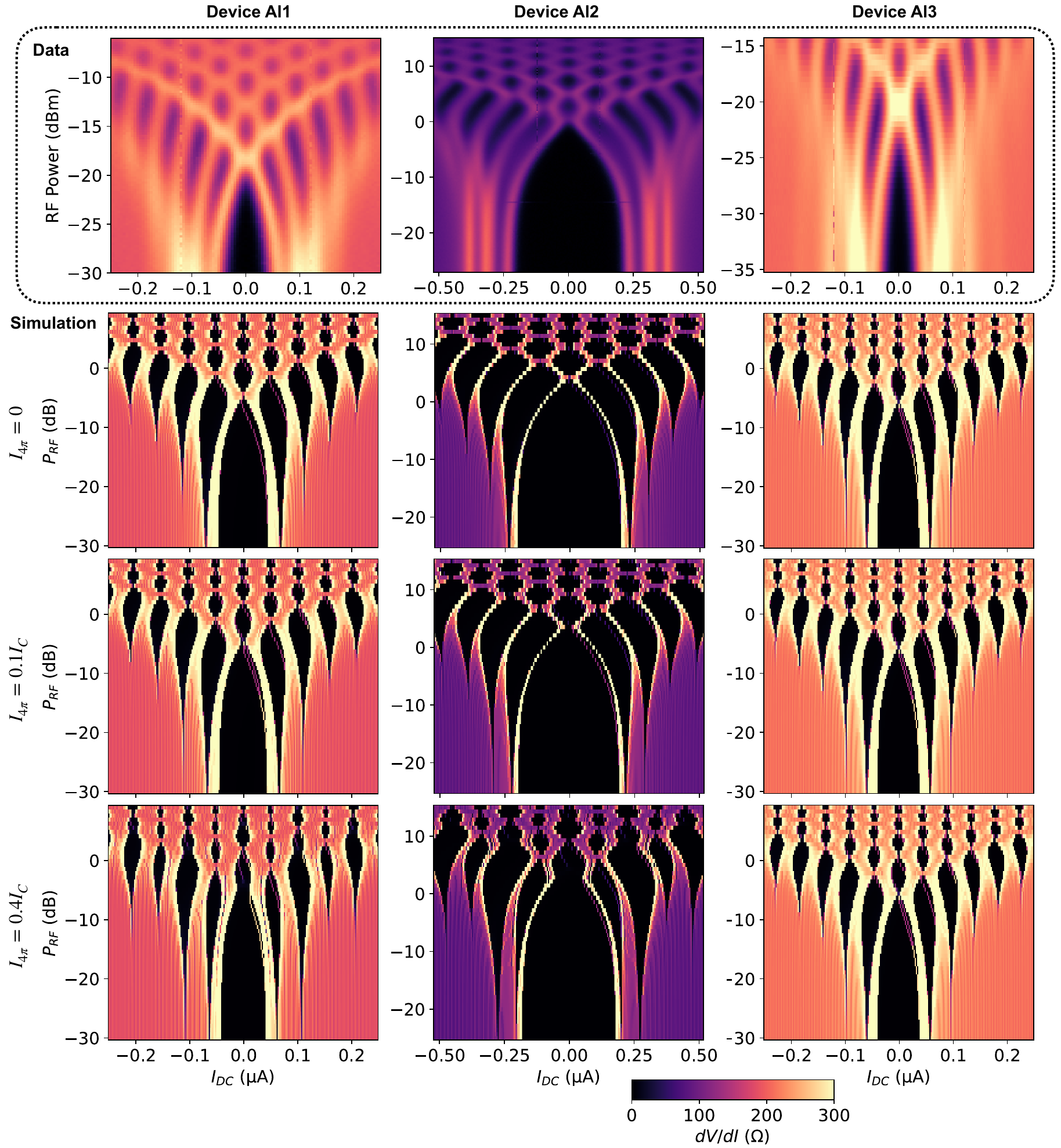}
	\caption{Simulated Shapiro steps in Devices~Al1, Al2, and Al3 at different ratios of the $4\pi$-periodic critical current. Experimental data is shown at top for reference.}
	\label{sfig:deviceAl123_sim}
\end{figure*}

\begin{figure*}
\centering
	\includegraphics[width=0.9\textwidth]{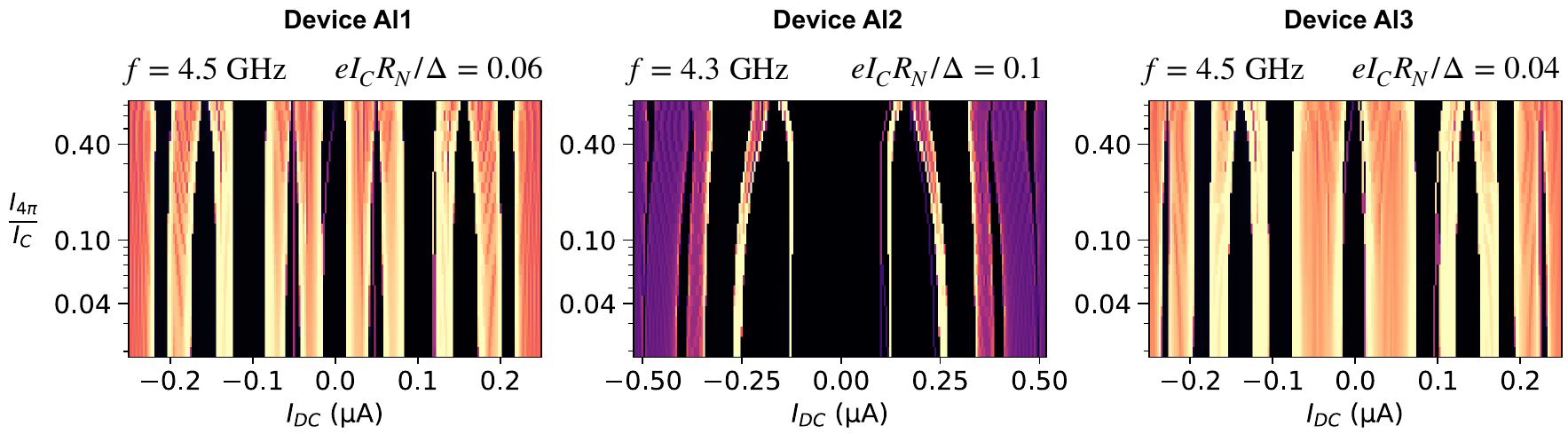}
	\caption{Simulated Shapiro steps in Devices~Al1-Al3 as the $4\pi$-periodic critical current is varied. Simulations use a constant power of $-2$~dB.}
	\label{sfig:sim_vs_4pi_Al}
\end{figure*}

Simulations of Devices~Al1, Al2, and Al3 are presented in Fig.~\ref{sfig:deviceAl123_sim}. The reduced frequency $\Omega=hf/2eI_C R_N$ is near unity (Device~Al1: $\Omega=1.1$ at 4.5~GHz; Device~Al2: $\Omega=0.55$ at 4.3~GHz; Device~Al3: $\Omega=1.1$ at 4.5~GHz). Therefore, the simulations feature broad transitions between Shapiro steps. Because Devices~Al1, Al2, and Al3 have much lower $I_C R_N$ products---$10\%$ to $20\%$ of that of Device~2---the suppression of odd Shapiro steps would be subtle even if the $4\pi$-periodic component were to constitutes a significant fraction of the total critical current. To summarize our findings, in Fig.~\ref{sfig:sim_vs_4pi_Al} we present simulations of the Shapiro steps as a function of the $I_{4\pi}$ components, clarifying the suppression of odd Shapiro steps is not readily visible in the devices with Al due to their smaller values of $I_C R_N$.

\clearpage

\section{Electron beam damage}

Electron beam lithography is widely used to pattern features at resolution better than optical wavelength. However electron beams damage \ce{Bi_2Se_3}-class materials. Beam damage has two quantifiable effects on mesoscopic electronic transport: a decrease in mobility $\mu$, due to added disorder, and a shift in the mean carrier density $n$, due to the addition of charged defects. Our results indicate that, to minimize damage, it is preferable to use low accelerating voltage for electron beam lithography. Writing at a higher voltage requires a larger electron dosage to expose resist (due to the decreased cross-section) and, in turn, more damage. Characterization and mitigation of damage to the BST film during electron beam lithography is discussed in detail in Ref.~\cite{andersen2023}. In the present work, we used PMMA resist, a 10~kV accelerating voltage, and an electron dose of 100~\textmu C/cm$^2$. At this irradiation dose, we did not observe significant changes to $n$ and $\mu$.

\section{Mean free path}

We estimate the inelastic mean free path of the BST film $\ell_i$ as the phase coherence length $\ell_\phi$, which can be studied through the weak anti-localization effect. We determine $\ell_\phi$ by fitting the magnetoconductance Hikami-Larkin-Nagaoka (HLN) formula

\begin{equation}
\sigma(B)-\sigma(B=0) \propto \Psi\left(\frac{1}{2} + \frac{B_0}{B}\right) - \ln\left(\frac{B_0}{B}\right)
\end{equation}

\begin{figure}[h]
\centering
	\includegraphics[width=0.4\textwidth]{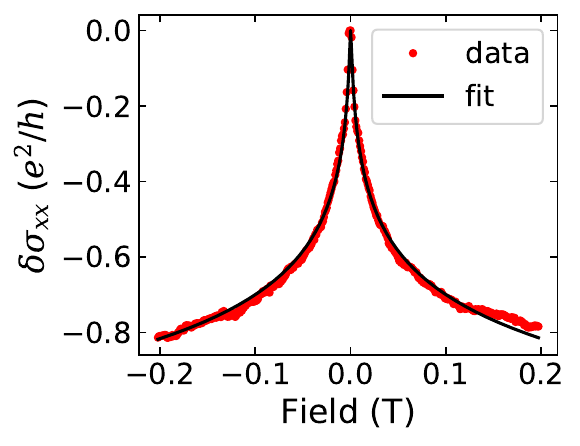}
	\caption{The magnetoconductance $\Delta\sigma_{xx}(B)=\sigma_{xx}(B)-\sigma_{xx}(B=0)$ of a Hall bar (located on the same chip as Devices~Al1-4) at 30~mK and a fit to the HLN formula. Data is symmetrized as $2\sigma_{xx}(B)=\sigma_{xx}^u(B)+\sigma_{xx}^d(B)$ where $\sigma_{xx}^{u(d)}$ is the conductance measured in an upwards (downwards) field sweep.}
	\label{sfig:WAL}
\end{figure}

Where $\Psi$ is the digamma function and $B_0 = \hbar/(4e\ell_\phi^2)$. The conductance of a Hall bar fashioned from the BST film alongside the HLN fit is shown in Fig.~\ref{sfig:WAL}. We extract $\ell_\phi = 850$~nm at 30~mK.

\newpage

\section{Engineering tunnel probes}

Tunnel spectroscopy is a common probe for Majorana zero modes (MZMs). MZMs should appear as a peak in the tunneling conductance at zero bias, although such an observation alone is insufficient to validate that the zero energy mode is indeed a MZM. Tunnel spectroscopy has been performed by scanning tunneling microscopy (STM) of vortices in unpatterned films of candidate intrinsic topological insulators and topological insulator-superconductor heterostructures~\cite{wang2018,zang2021}. Tunnel spectroscopy has also been performed in fabricated devices based on InAs and InSb in proximity with a superconductor~\cite{mourik2012}. The latter is possible because InAs and InSb are not intrinsically topological materials, and therefore have a band gap; a region tuned into the band gap by local gates serves as a tunnel barrier.

In this work we study fabricated devices based on an intrinsic topological insulator, BST. Fabricated devices are not well suited for STM because of surface adsorbates and contaminants imparted during ex situ device fabrication. Unlike the III-V material platforms, BST cannot be locally gated to form tunnel barriers because topological insulators have no full band gap: their topological nature demands the existence of surface states at energies inside the bulk bandgap. Tunnel spectroscopy of fabricated devices based on BST therefore demands the addition of a tertiary trivial insulator to serve as a tunnel barrier.

One approach to forming a tunnel barrier is to include a trivial insulating capping layer during MBE growth of the BST film. This approach mimics the \ce{CdTe} capping layer in \ce{CdTe}/\ce{HgTe} quantum wells and the \ce{In_{1-x}Al_xAs} capping layer often grown on \ce{InAs} 2DEGs, both of which have been used as tunnel barriers. However such a capping layer would be antagonistic to the contact transparency of a superconductor deposited later during device fabrication, which is needed to proximitize superconductivity in the topological insulator. An alternative approach is to deposit a thin trivial insulator after fabrication of the superconducting device. This approach is hindered by the difficulty of growing thin, high quality insulating layers in the low-temperature, plasma-free conditions required to avoid damaging the topological insulator. Our attempts at growing thin insulating oxides by atomic layer deposition were unsuccessful (despite our success in forming dielectric gates with thicker oxides). 

The ability of \ce{Pd} deposited ex situ to react with tellurides and form superconducting layers with high contact transparency offers the possibility of a new approach. Here, the MBE growth of BST is followed by MBE growth of a thin chalchogenide trivial insulator capping layer and evaporation of a metallic tunnel contact.  The trivial insulator could be \ce{CdTe_{1-x}Se_x} or \ce{(In_{1-x}Bi_x)_2(Te_{1-y}Se_y)_3}; epitaxial growth of \ce{CdSe}, \ce{(In_{1-x}Bi_x)_2Se_3}, and \ce{(In_{1-x}Sb_x)_2Te_3} layers on \ce{Bi_2Se_3}-class topological materials has been demonstrated~\cite{brahlek2016,jiang2018,Salehi2019}.

\begin{figure}[h]
\centering
	\includegraphics[width=0.6\textwidth]{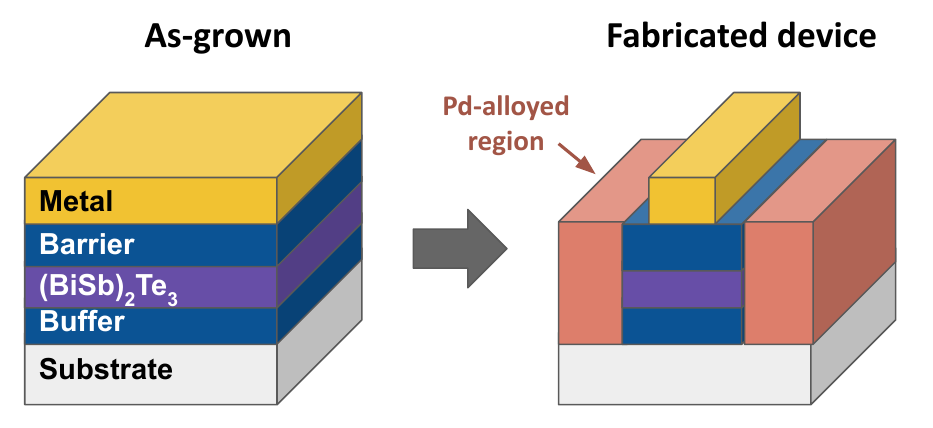}
	\caption{A diagram of a proposal for a device with tunnel contacts. Left, the heterostructure grown in an MBE chamber without breaking vacuum. Right, a finished device after ex situ fabrication. Crucially, the reaction between Pd and tellurides allows the self-formed superconductor to penetrate through the trivial insulating layer.}
	\label{sfig:junctionsketch}
\end{figure}

A single lithography step will serve as a mask for etching the metallic layer and depositing the superconductor. The capping layer protects the BST when the metallic tunnel contact is etched, for example, an \ce{Al} contact can be etched selectively against chalchogenides using commercially-available $\ce{H_3PO_4}$-based etchants. The undercut of the wet etch will prevent the tunnel contact from shorting to the superconductor. The superconductor will be formed by the deposition of \ce{Pd} (and perhaps \ce{Al}). The \ce{Pd} should react with the \ce{Te} throughout both the capping layer and the BST, forming a superconductor with high contact transparency to the BST. The remaining metal in the masked region serves as a contact to probe for zero-bias tunneling conductance features. Fig.~\ref{sfig:junctionsketch} shows a diagram of a device formed by this approach consisting of a Josephson junction whose weak link is interrogated by a tunnel contact~\cite{ren2019}.

\bibliography{references.bib}